\documentclass[12pt,twoside,a4paper]{article}
\pdfoutput=1

\usepackage{amsmath, amsthm, amsfonts, amssymb}
\usepackage{dsfont}
\usepackage{graphicx}
\usepackage{xcolor}
\usepackage[width=\textwidth,font=small,labelfont=bf,
labelsep=endash]{caption}
\usepackage[numbers,sort&compress]{natbib}

\normalsize
\long\def\@makecaption#1#2{%
  \vskip\abovecaptionskip
  \sbox\@tempboxa{\small #1. #2}%
  \ifdim \wd\@tempboxa >\hsize
    \small #1. #2\par
  \else
    \global \@minipagefalse
    \hb@xt@\hsize{\hfil\box\@tempboxa\hfil}%
  \fi
  \vskip\belowcaptionskip}

\newcommand\Tr            {\mathrm{Tr}}
\newcommand\ren           {R\'enyi\ }

\newcommand\doi[2]        {\href{http://dx.doi.org/#1}{#2}}

\renewcommand{\thesection}{\arabic{section}}
\renewcommand{\thesubsection}{\arabic{section}.\arabic{subsection}}
\renewcommand{\theequation}{\arabic{section}.\arabic{equation}}

\begin{document}

\title{An Inverse Mass Expansion for Entanglement Entropy in Free Massive Scalar Field Theory}
\author{Dimitrios Katsinis$^{1,2}$ and Georgios Pastras$^1$}
\date{$^1$NCSR ``Demokritos'', Institute of Nuclear and Particle Physics\\15310 Aghia Paraskevi, Attiki, Greece\\
$^2$National and Kapodistrian University of Athens, Department of Physics\\ 15784 Zografou, Attiki, Greece\linebreak\\
\texttt{pastras@inp.demokritos.gr dkatsinis@phys.uoa.gr}}

\vskip .5cm

\maketitle

\abstract{We extend the entanglement entropy calculation performed in the seminal paper by Srednicki \cite{Srednicki:1993im} for free real massive scalar field theories in $1+1$, $2+1$ and $3+1$ dimensions. We show that the inverse of the scalar field mass can be used as an expansion parameter for a perturbative calculation of the entanglement entropy. We perform the calculation for the ground state of the system and for a spherical entangling surface at third order in this expansion. The calculated entanglement entropy contains a leading area law term, as well as subleading terms that depend on the regularization scheme, as expected. Universal terms are non-perturbative effects in this approach. Interestingly, this perturbative expansion can be used to approximate the coefficient of the area law term, even in the case of a massless scalar field in $2+1$ and $3+1$ dimensions. The presented method provides the spectrum of the reduced density matrix as an intermediate result, which is an important advantage in comparison to the replica trick approach. Our perturbative expansion underlines the relation between the area law and the locality of the underlying field theory.}


\pagebreak

\setcounter{footnote}{0}

\def\thefootnote{\arabic{footnote}}

\newpage
\tableofcontents

\setcounter{equation}{0}
\section{Introduction}
\label{sec:intro}
Quantum entanglement is the physical phenomenon that appears when a composite quantum system lies in a state such that no description of the state of its subsystems is available. In the presence of quantum entanglement, measurements in the entangled subsystems are correlated. The most well known example of an entangled system, the so called EPR paradox \cite{Einstein:1935rr}, requires just two spinors; it was initially conceived as contradictory to causality, and, thus, as an adequate theoretical experiment to question the completeness of the quantum description of nature. However, later on, the corresponding correlations were verified experimentally.

A quantum subsystem $A$ entangled to its environment $A^C$ cannot be described by a state; it is rather described by a density matrix $\rho_A$, calculable by tracing out the degrees of freedom of the subsystem $A^C$ from the overall density matrix $\rho$
\begin{equation}
\rho_A  = \Tr_{{A}^c} \rho .
\end{equation}
In the absence of entanglement, there is a state description for the subsystem $A$, and, thus, this reduced density matrix $\rho_A$ corresponds to a pure state; on the contrary, in the case entanglement is present, the reduced density matrix corresponds to a mixed state. The above indicate that the entanglement is encoded in the spectrum of the reduced density matrix $\rho_A$. It follows that a natural choice for a measure of entanglement is Shannon entropy applied to the spectrum of $\rho_A$, known as Entanglement Entropy, $S_{\textrm{EE}}$,
\begin{equation}
S_{\textrm{EE}}  :=  - \Tr \left( {\rho_A \ln \rho _A } \right) .
\label{eq:entanglement_entropy_definition}
\end{equation}
Entanglement entropy has found a large variety of applications to many physics sectors including quantum computing \cite{Abramsky:2003,Abramsky:2004,Abramsky:2005,Abramsky:2008,Coecke:2010,Baez:2011a,Vicary:2012,Baez:2014}, condensed matter systems \cite{Levin:2006zz,Kitaev:2005dm,Hamma:2005,Calabrese:2004eu,Calabrese:2005zw}, as well as quantum gravity and the holographic duality \cite{Ryu:2006bv,Ryu:2006ef,Nishioka:2009un,Takayanagi:2012kg,VanRaamsdonk:2009ar,VanRaamsdonk:2010pw,Bianchi:2012ev,Myers:2013lva,Balasubramanian:2013rqa}.

In a seminal paper \cite{Srednicki:1993im}, Srednicki performed a numerical calculation of entanglement entropy for a real free massless scalar field theory at its ground state, considering as subsystem $A$ the degrees of freedom inside a sphere of radius $R$. The surprising result shows that entanglement entropy is not proportional to the volume of the sphere, but rather to its area. This profound similarity to the black hole entropy \cite{Bekenstein:1973ur,Bardeen:1973gs,Hawking:1974sw}, discussed even before Srednicki's calculation \cite{Bombelli:1986rw}, became even more intriguing after the development of the holographic dualities \cite{Maldacena:1997re,Gubser:1998bc,Witten:1998qj} and the Ryu-Takayanagi conjecture \cite{Ryu:2006bv,Ryu:2006ef}, which interrelates entanglement entropy in the boundary conformal field theory to the geometry of the bulk. The latter may allow the perspective of understanding the black hole entropy as entanglement entropy, and the gravitational interactions as an entropic force associated with quantum entanglement statistics \cite{ Verlinde:2010hp,Lashkari:2013koa,Faulkner:2013ica,Bakas:2015opa}.

In this context, the further investigation of the similarities between gravitational and quantum entanglement physics and the development of appropriate tools for their study presents a certain interest. In this work, we extend the original entanglement entropy calculation presented in \cite{Srednicki:1993im} to massive free scalar field theory and develop a perturbative method for the calculation of entanglement entropy in such systems.

The majority of entanglement entropy calculations in field theory are based on the replica trick \cite{Calabrese:2004eu,Callan:1994py,Holzhey:1994we,Casini:2009sr,Hertzberg:2010uv,Liu:2012eea,Lewkowycz:2012qr}. This technique is based on the calculation of the entanglement \ren entropies $S_q$ for an arbitrary positive integer index $q>1$. (The entanglement \ren entropy $S_q$ is defined as $S_q := \frac{1}{{1 - q}}\ln \Tr \rho_A ^q$.) Then, the entanglement entropy is recovered from the analytic continuation of $S_q$ as the limit $S_{\mathrm{EE}} = \mathop {\lim }\limits_{q \to 1} S_q$. Although the entanglement \ren entropies $S_q$ in principle contain the whole information of the reduced density matrix spectrum, the process of deriving the latter from the former is complicated. Relevant calculations are usually restricted to the specification of the largest eigenvalue and its degeneracy. Furthermore, holographic calculations cannot provide more information on the reduced density matrix spectrum other than the entanglement entropy, due to the very nature of the Ryu-Takayanagi conjecture \cite{Ryu:2006bv,Ryu:2006ef}. An important feature of Srednicki's calculation is the fact that it is not limited to the calculation of entanglement entropy; on the contrary the full spectrum of $\rho_A$ is an intermediate result. As we discussed above, quantum entanglement is encoded into the spectrum of $\rho_A$; the entanglement entropy is just one piece of information. Therefore, although they are old, the methods of \cite{Srednicki:1993im} present a certain advantage.

The structure of this paper is as follows: in Section \ref{sec:qft}, we review the derivation of entanglement entropy in systems of coupled harmonic oscillators lying at their ground state and extend the calculation in free scalar field theory including a mass term, closely following \cite{Srednicki:1993im}. In section \ref{sec:expansion}, we show that the inverse of the scalar field mass can be used as an expansion parameter allowing a perturbative calculation of entanglement entropy and develop the basic formulae of this perturbation theory. In section \ref{sec:area}, we perform the perturbative calculation for massive free scalar field theory in $1+1$, $2+1$ and $3+1$ dimensions and show that the leading contribution to the entanglement entropy for large entangling sphere radii obeys an area law; we specify the relevant coefficients and the first subleading corrections and we compare with numerical calculations. In Section \ref{sec:discussion}, we discuss our results. Appendix \ref{sec:descretization} contains the details of Srednicki's regularization scheme. Appendix \ref{sec:third} contains the details of the perturbative calculation of entanglement entropy at second and third order. Finally, appendix \ref{sec:code} contains the code used for the numerical calculations of entanglement entropy.

\section{Entanglement Entropy in Free Scalar QFT}
\label{sec:qft}

\subsection{Entanglement Entropy of Coupled Oscillators}
\label{subsec:qm}
The first step towards the calculation of entanglement entropy in free scalar field theory is the calculation of the latter in a system of coupled harmonic oscillators at its ground state. This problem has been solved long ago; here, we briefly sketch its solution. More details are provided in \cite{Srednicki:1993im}.

Assume a system of $N$ coupled harmonic oscillators described by the quadratic Hamiltonian
\begin{equation}
H = \frac{1}{2}\sum\limits_{i = 1}^N {p_i^2}  + \frac{1}{2}\sum\limits_{i,j = 1}^N {{x_i}{K_{ij}}{x_j}} ,
\end{equation}
where the matrix $K$ is symmetric and has positive eigenvalues, as required for the vacuum stability. Since $K$ has been positively defined, its square root $\Omega := \sqrt{K}$ can be appropriately defined, so that it also has positive eigenvalues.

In the following, without loss of generality, the subsystem $A$ is considered to comprise of $N-n$ oscillators, those described by coordinates $x_i$ with $i > n$. It follows that its complementary subsystem $A^C$ comprises of the $n$ oscillators described by coordinates $x_i$ with $i \le n$. We may write the matrix $\Omega$ in block form as
\begin{equation}
\Omega  = \left( {\begin{array}{*{20}{c}}
A&B\\
{{B^T}}&C
\end{array}} \right) , \label{eq:omega_block}
\end{equation}
where $A$ is an $n\times n$, $C$ is an $\left(N-n\right)\times \left(N-n\right)$ and $B$ is an $n\times \left(N-n\right)$ matrix.

We define the $\left(N-n\right)\times \left(N-n\right)$ matrices $\beta$ and $\gamma$ as,
\begin{align}
\beta &:= \frac{1}{2}{B^T}{A^{ - 1}}B, \label{eq:beta_def}\\
\gamma &:= C - \frac{1}{2}{B^T}{A^{ - 1}}B = C - \beta. \label{eq:gamma_def}
\end{align}
Let $\lambda_{i}$, $i = n + 1, \ldots ,N$, be the eigenvalues of the matrix $\gamma^{-1} \beta$. Then, the spectrum of the reduced density matrix $\rho_A$ is given by
\begin{equation}
{p_{{n_{n + 1}}, \ldots ,{n_N}}} = \prod\limits_{i = n + 1}^N {\left( {1 - {\xi _i}} \right)\xi _i^{{n_i}}} ,\quad n_i \in \mathbb{Z},
\label{eq:several_eigenvalues}
\end{equation}
where
\begin{equation}
{\xi _i} = \frac{{{\lambda_{i}}}}{{1 + \sqrt {1 - \lambda_{i}^2} }} .
\end{equation}
It follows that the entanglement entropy is given by
\begin{equation}
S_{\textrm{EE}} \left( {N,n} \right) = \sum\limits_{j = n + 1}^N {\left( { - \ln \left( {1 - {\xi _j}} \right) - \frac{{{\xi _j}}}{{1 - {\xi _j}}}\ln {\xi _j}} \right)} .
\end{equation}

The ground state of a system of coupled harmonic oscillators is a highly entangled state. The specification of entanglement entropy at the ground state requires a non-trivial, non-perturbative calculation. However, there is a small window allowing a perturbative approach. By definition, the matrix $B$ does not contain any of the diagonal elements of the matrix $\Omega = \sqrt{K}$. Therefore, the matrix $\beta$, and, thus, the eigenvalues of $\gamma^{-1}\beta$, as well as the entanglement entropy, can be perturbatively calculated, in the case the diagonal elements of the matrix $K$ are much larger than its non-diagonal elements. As the non-diagonal elements of $K$ describe the couplings between the harmonic oscillators, in such an expansion, the zero-th order result is the entanglement entropy in a system of decoupled oscillators at their ground state, i.e. vanishing entanglement entropy.

The entanglement entropy is a valuable measure of entanglement, however, it does not contain the whole information. The latter is contained in the full spectrum of the reduced density matrix $\rho_A$. An important advantage of the approach followed in this work is the fact that it allows the direct calculation of the latter through equation \eqref{eq:several_eigenvalues}, as an intermediate step towards the calculation of entanglement entropy.

\subsection{Entanglement Entropy in Free Scalar Field Theory}
\label{subsec:massless}
In the approach of \cite{Srednicki:1993im}, the degrees of freedom of the scalar field theory are discretized
via the introduction of a lattice of spherical shells, and, thus, the introduction of a UV cutoff. Furthermore, an IR cutoff is imposed, putting the system in a spherical box. This inhomogeneous distretization may appear disadvantageous, as it breaks some of the symmetries of the theory; although it preserves rotations, it breaks boosts and translations. However, the consideration of the stationary entangling sphere, which separates the degrees of freedom to two subsystems, has already broken these symmetries. This approach reduces the problem of the calculation of entanglement entropy in field theory to a similar quantum mechanics problem with finite degrees of freedom. Since we are studying free scalar field theory, the latter quantum mechanical system is simply a system of coupled oscillators with a quadratic Hamiltonian at its ground state. More details on this discretazation scheme are provided in appendix \ref{sec:descretization}.

\subsubsection*{$3+1$ Dimensions}
Let us consider a free real scalar field theory in $3+1$ dimensions. The Hamiltonian equals
\begin{equation}
H = \frac{1}{2}\int {{d^3}x\left[ {{\pi ^2}\left( {\vec x} \right) + {{\left| {\vec \nabla \varphi \left( {\vec x} \right)} \right|}^2} + {\mu^2}{\varphi^2} {{\left( {\vec x} \right)}}} \right]} .
\label{eq:hamiltonian_3}
\end{equation}
Decomposing the field to real spherical harmonics ${Y_{\ell m}}$, we find that the corresponding components ${\varphi _{\ell m}}\left( r \right)$ obey canonical commutation relations of the form
\begin{equation}
\left[ {{\varphi _{\ell m}}\left( r \right),{\pi _{\ell ' m'}}\left( r' \right)} \right] = i\delta \left( {r - r'} \right){\delta _{\ell \ell '}}{\delta _{mm'}} ,
\end{equation}
where $r=\left|\vec{x}\right|$ is the radial coordinate. 

The only continuous variable left is the radial coordinate $r$. We regularize the theory introducing a lattice of $N$ spherical shells with radii $r_i = i a$ with $i \in \mathds{N}$ and $1 \le i \le N$. The radial distance between consequent spherical shells introduces a UV cutoff $1/a$, while the overall size of the lattice imposes an IR cutoff $1/({Na})$. The introduction of the spherical lattice sets the number of degrees of freedom for each pair $\left( \ell , m \right)$ finite. The discretized Hamiltonian reads
\begin{equation}
H = \frac{1}{{2a}}\sum\limits_{\ell ,m} {\sum\limits_{j = 1}^N {\left[ {\pi _{\ell m,j}^2 + {{\left( {j + \frac{1}{2}} \right)}^2}{{\left( {\frac{{{\varphi _{\ell m,j + 1}}}}{{j + 1}} - \frac{{{\varphi _{\ell m,j}}}}{j}} \right)}^2} + \left( {\frac{{\ell \left( {\ell  + 1} \right)}}{{{j^2}}} + {\mu^2}{a^2}} \right)\varphi _{\ell m,j}^2} \right]} } .
\end{equation}

Different $\ell $ and $m$ indices do not mix and furthermore the index $m$ does not appear explicitly in the Hamiltonian. It follows that the problem can be split to infinite independent sectors, identified by the index $\ell $, each containing $2 \ell + 1$ identical subsectors. We consider an entangling sphere of radius $R=\left(n+1/2\right) a$. Then, the entanglement entropy at the ground state is given by
\begin{equation}
S_{\textrm{EE}}\left( {N,n} \right) = \sum\limits_{\ell=0}^{\infty}  {\left( {2\ell  + 1} \right){S_{\ell}}\left( {N,n} \right)} ,
\label{eq:qft_entropy_series}
\end{equation}
where ${S_{\ell }}\left( {N,n} \right)$ is the entanglement entropy corresponding to the ground state of the Hamiltonian
\begin{equation}
H_\ell = \frac{1}{{2a}} {\sum\limits_{j = 1}^N {\left[ {\pi _{\ell ,j}^2 + {{\left( {j + \frac{1}{2}} \right)}^2}{{\left( {\frac{{{\varphi _{\ell ,j + 1}}}}{{j + 1}} - \frac{{{\varphi _{\ell ,j}}}}{j}} \right)}^2} + \left( {\frac{{\ell \left( {\ell  + 1} \right)}}{{{j^2}}} + {\mu^2}{a^2}} \right)\varphi _{\ell ,j}^2} \right]} } .
\label{eq:qft_hamiltonian_l}
\end{equation}
The quadratic Hamiltonian \eqref{eq:qft_hamiltonian_l} describes $N$ harmonically coupled oscillators, and, thus, the problem of the calculation of ${S_{\ell }}\left( {N,n} \right)$ has been reduced to the class of problems solved in section \ref{subsec:qm}.

For large $\ell$, the Hamiltonian $H_\ell$ becomes almost diagonal. Therefore, for large $\ell$, the degrees of freedom are almost decoupled, and, thus, the system \eqref{eq:qft_hamiltonian_l} at its ground state is almost disentangled. It can be shown that ${S_{\ell }}\left( {N,n} \right)$ decreases with $\ell$ fast enough so that the series \eqref{eq:qft_entropy_series} is converging \cite{Srednicki:1993im,Riera:2006vj}.

\subsubsection*{$2+1$ Dimensions}

In a similar manner, we may study free scalar field theory in $2+1$ dimensions. The Hamiltonian reads
\begin{equation}
H = \frac{1}{2}\int {{d^2}x\left[ {{\pi ^2}\left( {\vec x} \right) + {{\left| {\vec \nabla \varphi \left( {\vec x} \right)} \right|}^2} + {\mu^2}{\varphi^2} {{\left( {\vec x} \right)}}} \right]} .
\label{eq:hamiltonian_2}
\end{equation}
We expand the field to real circular harmonics and then introduce a lattice of circular shells to find the discretized Hamiltonian
\begin{equation}
H = \frac{1}{{2a}}\sum\limits_{\ell} {\sum\limits_{j = 1}^N {\left[ {\pi _{\ell,j}^2 + {{\left( {j + \frac{1}{2}} \right)}}{{\left( {\frac{{{\varphi _{\ell ,j + 1}}}}{{\sqrt{j + 1}}} - \frac{{{\varphi _{\ell ,j}}}}{\sqrt{j}}} \right)}^2} + \left( {\frac{{\ell^2 }}{{{j^2}}} + {\mu^2}{a^2}} \right)\varphi _{\ell ,j}^2} \right]} } .
\end{equation}

Different $\ell $ indices do not mix. Therefore, in a similar manner to the problem at $3+1$ dimensions, the problem can be split to infinite independent sectors, identified by the index $\ell $. The entanglement entropy at the ground state is given by
\begin{equation}
S_{\textrm{EE}}\left( {N,n} \right) = \sum\limits_{\ell=-\infty}^{\infty}  { {S_{\ell}}\left( {N,n} \right)} ,
\label{eq:qft_entropy_series_d2}
\end{equation}
where ${S_{\ell }}\left( {N,n} \right)$ is the entanglement entropy corresponding to the ground state of the Hamiltonian
\begin{equation}
H_\ell = \frac{1}{{2a}} {\sum\limits_{j = 1}^N {\left[ {\pi _{\ell ,j}^2 + {{\left( {j + \frac{1}{2}} \right)}}{{\left( {\frac{{{\varphi _{\ell ,j + 1}}}}{{\sqrt{j + 1}}} - \frac{{{\varphi _{\ell ,j}}}}{\sqrt{j}}} \right)}^2} + \left( {\frac{{\ell^2 }}{{{j^2}}} + {\mu^2}{a^2}} \right)\varphi _{\ell ,j}^2} \right]} } .
\label{eq:qft_hamiltonian_l_d2}
\end{equation}
The calculation of the latter lies within the class of problems solved in section \ref{subsec:qm}.

\subsubsection*{$1+1$ Dimensions}
Finally, we consider a free real scalar field theory in $1+1$ dimensions. The Hamiltonian reads
\begin{equation}
H = \frac{1}{2}\int {{d}x\left[ {{\pi ^2}\left( {x} \right) + {{\left| { \frac{\partial}{\partial x} \varphi \left( {x} \right)} \right|}^2} + {\mu^2}{\varphi^2} {{\left( {x} \right)}}} \right]} .
\label{eq:discretize_hamiltonian_d1}
\end{equation}
We may directly apply the same discretization scheme to obtain
\begin{equation}
H = \frac{1}{{2a}} \sum\limits_{j = 1}^N {\left[ {\pi _{\ell,j}^2 + {{\left( \varphi _{\ell ,j + 1}- \varphi _{j}\right)}^2} + {\mu^2}{a^2}\varphi _{j}^2} \right]} .
\label{eq:qft_hamiltonian_l_d1}
\end{equation}

\section{An Inverse Mass Expansion for Entanglement Entropy}
\label{sec:expansion}

The discretized Hamiltonians \eqref{eq:qft_hamiltonian_l}, \eqref{eq:qft_hamiltonian_l_d2} and \eqref{eq:qft_hamiltonian_l_d1} are describing a system of $N$ coupled harmonic oscillators that falls within the class of systems studied in section \ref{subsec:qm}. We may thus proceed to calculate the entanglement entropy following the scheme of this section.

\subsection{An Inverse Mass Expansion}
\label{subsec:massive}

As an indicative example, in $3+1$ dimensions, the $K$ matrix describing the interactions between the harmonic oscillators can be directly read from equation \eqref{eq:qft_hamiltonian_l},
\begin{multline}
{K_{ij}} = \left( {{{\left( {\frac{{i + \frac{1}{2}}}{i}} \right)}^2} + {{\left( {\frac{{i - \frac{1}{2}}}{i}} \right)}^2} \left( 1- \delta_{i1} \right) + \frac{{l\left( {l + 1} \right)}}{{{i^2}}} + {\mu^2}{a^2}} \right){\delta _{ij}} \\
- \frac{{{{\left( {i + \frac{1}{2}} \right)}^2}}}{{i\left( {i + 1} \right)}}{\delta _{i + 1,j}} - \frac{{{{\left( {j + \frac{1}{2}} \right)}^2}}}{{j\left( {j + 1} \right)}}{\delta _{i,j + 1}} ,
\end{multline}
where $i,j = 1,2, \ldots ,N$. As we have commented in section \ref{subsec:qm}, a perturbation theory can be applied when the diagonal elements of the matrix $K$ are much larger than the non-diagonal ones. This criterion clearly is satisfied at the limit of a very large mass $\mu$. A similar approach is followed in \cite{Riera:2006vj} focusing in the behaviour of entanglemment entropy for large $\ell$. It has to be pointed out that the actual expansion parameter is neither $m$ nor $\ell$, but the diagonal elements of $K$ themselves.

In all numbers of dimensions under study, the matrix $K$ is of the form
\begin{equation}
{K_{ij}} = K_i{\delta _{ij}} + L_i \left({\delta _{i + 1,j}} +{\delta _{i,j + 1}} \right).
\label{eq:K_ij_def}
\end{equation}
We define the quantities $k_i$ and $l_i$ so that
\begin{align}
K_i &:= \frac{k_i^2}{\varepsilon^2} , \label{eq:k_i_def}\\
L_i &:= l_i \left(k_i + k_{i+1} \right) . \label{eq:l_i_def}
\end{align}
The parameter $\varepsilon$ is the expansion parameter of the perturbation theory that we are about to develop, which is obviously of order $1/\mu$. The expansion in $\varepsilon$ is also a semiclassical expansion; recovering the fundamental constants in the dimensionless expansion parameter $\varepsilon$, the latter assumes the form $\hbar / \left( \mu a c \right)$. This is in line with the fact that the zeroth order entanglement entropy in this perturbative approach vanishes.

In order to calculate the desired entanglement entropy, we need to calculate the square root $\Omega$ of the matrix $K$, then the matrices $\beta$, $\gamma$ and finally the eigenvalues of $\gamma^{-1}\beta$, perturbatively in $\varepsilon$. There is one important detail that has to be taken into account in these perturbative calculations. Since the lowest order elements of $K$ are the diagonal ones, this is also going to be the case for its square root $\Omega$. However, the matrix $B$, being an off-diagonal block of the matrix $\Omega$, does not contain such elements. The lowest order elements of $B$ are the first subleading elements that appear in $\Omega$. As a result, preserving a certain order in perturbation theory requires the calculation of the square root of $K$ at one order higher than the desired order. In the following, we will present the calculation at first non-vanishing order, therefore we will keep two non-vanishing terms in the expansion of $\Omega$.

The square root of the matrix $K$, with two non-vanishing terms equals
\begin{equation}
{\Omega_{ij}} = k_i \delta _{ij} {\varepsilon}^{-1} + {l_i} \left({\delta _{i + 1,j}} +{\delta _{i,j + 1}} \right) \varepsilon + \mathcal{O} \left( \varepsilon^3\right) .
\end{equation}
The blocks $A$, $B$ and $C$ of the matrix $\Omega$ obviously equal
\begin{align}
{A_{ij}} &= {k_i}{\delta _{ij}}{\varepsilon }^{-1} + {l_i}\left( {{\delta _{i + 1,j}} + {\delta _{i,j + 1}}} \right)\varepsilon  + \mathcal{O}\left( {{\varepsilon ^3}} \right) , \\
{B_{ij}} &= {l_n}{\delta _{i,n}}{\delta _{j,1}}\varepsilon  + \mathcal{O}\left( {{\varepsilon ^3}} \right) , \\
{C_{ij}} &= {k_{i + n}}{\delta _{ij}}{\varepsilon }^{-1} + {l_{i + n}}\left( {{\delta _{i + 1,j}} + {\delta _{i,j + 1}}} \right)\varepsilon  + \mathcal{O}\left( {{\varepsilon ^3}} \right) .
\end{align}
It is noteworthy that the above formulae contain only odd powers of $\varepsilon$. Furthermore, the matrix $B$ contains only order $\varepsilon$ terms to this order, as it does not contain any diagonal elements of $\Omega$. Had we desired to calculate the eigenvalues of the reduced density matrix with two non-vanishing terms in the $\varepsilon$ expansion, we should have calculated $\Omega$ at $\varepsilon^5$ order.

From now on, we need to keep only one non-vanishing term in our expressions. The matrices $A^{-1}$ and $C^{-1}$ equal
\begin{align}
{\left(A^{-1}\right)_{ij}} &= \frac{1}{k_i}{\delta _{ij}}{\varepsilon } + \mathcal{O}\left( {{\varepsilon ^3}} \right) , \\
{\left(C^{-1}\right)_{ij}} &= \frac{1}{k_{i + n}}{\delta _{ij}}{\varepsilon } + \mathcal{O}\left( {{\varepsilon ^3}} \right) .
\end{align}
The matrix $\gamma^{-1}$ is identical to the matrix $C^{-1}$ at this order. The matrix $\beta$ has a single non-vanishing element at this order, namely,
\begin{equation}
{\beta _{ij}} = \frac{l_n^2}{{{2k_n}}}{\delta _{i,1}}{\delta _{j,1}}{\varepsilon ^3} + \mathcal{O}\left( {{\varepsilon ^5}} \right) .
\end{equation}
Finally, 
\begin{equation}
{\left( {{\gamma ^{ - 1}}\beta } \right)_{ij}} = \frac{l_n^2}{{{2 k_n k_{n + 1}}}}{\delta _{i,1}}{\delta _{j,1}}{\varepsilon ^4} + \mathcal{O}\left( {{\varepsilon ^6}} \right) .
\end{equation}

Obviously, the matrix ${{\gamma ^{ - 1}}\beta }$ has only one non-vanishing eigenvalue at this order, being equal to its sole non-vanishing element,
\begin{align}
{\lambda _{1}} &= \frac{l_n^2}{{{2 k_n k_{n + 1}}}} {\varepsilon ^4} + \mathcal{O}\left( {{\varepsilon ^8}} \right), \\
{\lambda _{i}} &= \mathcal{O}\left( {{\varepsilon ^8}} \right),\quad i > 1 .
\end{align}

Thus, the entanglement entropy at first non-vanishing order equals
\begin{equation}
{S_{{\rm{EE}\ell}}} = \frac{l_n^2}{{4{k_n}{k_{n + 1}}}}\left( {1 - \ln \frac{l_n^2 {\varepsilon ^4}}{{4{k_n}{k_{n + 1}}}} } \right) {\varepsilon ^4} + \mathcal{O}\left( {{\varepsilon ^8}} \right) .
\label{eq:SEEl_first_order}
\end{equation}

\subsection{The Expansion at Higher Orders}
\label{subsec:higher_orders}

Continuing the expansion at higher orders, several patterns appear in the form of the expansions of the related matrices. More specifically, as long as the matrix $\Omega$ is considered:
\begin{itemize}
\item Only odd powers of $\varepsilon$ appear in the expansion of $\Omega$.
\item The leading term in any element in the $k$-diagonal is of order $\varepsilon^{2k-1}$. Therefore, the matrix $\Omega$ calculated with $n$ non-vanishing terms in the perturbation theory contains non-vanishing elements up to the $\left(n-1\right)$-diagonal.
\item Any subleading term in the elements of the matrix $\Omega$ is four orders higher than the previous one. Thus, an element in the $k$-diagonal is written as a series of the form
\begin{equation}
{\Omega _{i,i + k}} = {\Omega _{i + k,i}} = \sum\limits_{n = 0}^\infty  {\omega _i^{k\left( {2k - 1 + 4n} \right)}{\varepsilon ^{2k - 1 + 4n}}} , \quad i = 1, \ldots , N - k .
\end{equation}
\end{itemize}
We use the above notation with three indices for the coefficients of the expansion. The subscript denotes the line number if the element lies on the top triangle of the matrix or the column number if it lies in the bottom triangle, the superscript denotes the number of the diagonal, whereas the superscript in parentheses is the order of the term in the $\varepsilon$ expansion. The matrices $A^{-1}$ and $C^{-1}$ follow the same pattern with an overall increase by 2 to all orders. Namely,
\begin{align}
{\left( A^{-1} \right) _{i,i + k}} &= {\left( A^{-1} \right) _{i + k,i}} = \sum\limits_{n = 0}^\infty  {a _i^{k\left( {2k + 1 + 4n} \right)}{\varepsilon ^{2k + 1 + 4n}}} , \quad i = 1, \ldots , n - k , \\
{\left( C^{-1} \right) _{i,i + k}} &= {\left( C^{-1} \right) _{i + k,i}} = \sum\limits_{n = 0}^\infty  {c _i^{k\left( {2k + 1 + 4n} \right)}{\varepsilon ^{2k + 1 + 4n}}} , \quad i = 1, \ldots , N - n - k .
\end{align}

The matrix $\gamma$ is defined as $\gamma  = C - \beta $. The expansion for $\gamma^{-1} \beta$ can be acquired using the formula
\begin{equation}
{\gamma ^{ - 1}}\beta  = \sum\limits_{n = 1}^\infty  {{{\left( {{C^{ - 1}}\beta } \right)}^n}} .
\end{equation}
The form of the expansions of $\Omega$, $A^{-1}$ and $C^{-1}$ imply that the expansion of the matrix $\gamma^{-1} \beta$, whose eigenvalues define the spectrum of the reduced density matrix, follows a similar pattern. In this case, the leading order element is the $\left( 1,1 \right)$ element, which is of order $\varepsilon^4$. Every offset by a column or a row increases the order of the leading term by 2. Again subleading terms in any element are four orders higher than the previous one,
\begin{equation}
{\left( \gamma^{-1} \beta \right) _{ij}} = \sum\limits_{n = 0}^\infty  {\beta _{ij}^{\left( {2i+2j+4n} \right)}{\varepsilon ^{2i+2j+4n}}} .
\end{equation}
A direct consequence of the above is the fact that the eigenvalues of $\gamma^{-1} \beta$ are naively expected to have a given hierarchy. The largest eigenvalue is of order $\varepsilon^4$, the second largest is of order $\varepsilon^8$ and so on.

The calculation at the next to the leading order is analytically presented in appendix \ref{sec:third}. It turns out that the second largest eigenvalue vanishes at this order, whereas the largest eigenvalue receives corrections at order $\varepsilon^8$. At third order the calculation is straightforward but more messy. The result is presented in the appendix only in the appropriate limit for the specification of the ``area law'' contribution to the entanglement entropy that we will discuss in next section. At this order, the largest eigenvalue receives another correction at order $\varepsilon^{12}$, while one more non-vanishing eigenvalue emerges, with a leading contribution at the same order. As a general rule, a new non-vanishing eigenvalue emerges every second order in the perturbation theory. The corrections to the largest eigenvalue at a given order in the expansion have a more important effect to the entanglement entropy than the emergence of new eigenvalues at the same order.

\section{Area and Entanglement in the Inverse Mass Expansion}
\label{sec:area}
\subsection{The Leading ``Area Law'' Term}
\label{subsec:area_law}

In section \ref{sec:expansion} we managed to acquire an expansive formula for entanglement entropy. In order to study the dependence of entanglement entropy on the size of the entangling sphere, we need to expand our results for large entangling sphere radii. We assume that the entangling sphere lies exactly in the middle between the $n$-th and $\left(n+1\right)$-th site of the spherical lattice. We define
\begin{equation}
n_R := n+\frac{1}{2},
\end{equation}
so that $R=n_R a$ is the radius of the entangling sphere. In the following, we will study the expansion of entanglement entropy for large $n_R$.

\subsubsection*{$3+1$ Dimensions}

In $3+1$ dimensions, the entanglement entropy equals the sum of entanglement entropy from all $\ell$ sectors, as shown in equation \eqref{eq:qft_entropy_series}. The summation of this series cannot be performed analytically. For this reason, we will use the Euler-Maclaurin formula
\begin{multline}
\sum\limits_{n = a}^b {f\left( n \right)}  = \int_a^b {dxf\left( x \right)}  + \frac{{f\left( a \right) + f\left( b \right)}}{2}\\
 + \sum\limits_{k = 1}^\infty  {\frac{{{B_{2k}}}}{{\left( {2k} \right)!}}\Bigg[ {{{\left. {\frac{{{d^{2k - 1}}f\left( x \right)}}{{d{x^{2k - 1}}}}} \right|}_{x = b}} - {{\left. {\frac{{{d^{2k - 1}}f\left( x \right)}}{{d{x^{2k - 1}}}}} \right|}_{x = a}}} \Bigg]} ,
\label{eq:Euler}
\end{multline}
to approximate the series by the integral
\begin{equation}
{S_{{\rm{EE}}}} = \sum\limits_{\ell  = 0}^\infty  {\left( {2\ell  + 1} \right){S_{\ell }}\left( {N,n} \right)}  \simeq \int_0^\infty  {d\ell \left( {2\ell  + 1} \right){S_{\ell }}\left( {N,n,{{\ell \left( {\ell  + 1} \right)}}} \right)} .
\end{equation}
The coefficients $B_k$ are the Bernoulli numbers defined so that $B_1 = 1 / 2$.

We would like to expand the above integral for large $R$. This cannot be done directly, since $n_R$ appears in $S_{\ell }$ in the form of the fraction ${{\ell \left( {\ell  + 1} \right)}}/{{{n_R^2}}}$ and $\ell$ becomes arbitrarily large within the integration range. This problem may be bypassed performing the change of variables ${{\ell \left( {\ell  + 1} \right)}}/{{{n_R^2}}} = y$ to find
\begin{equation}
{S_{{\rm{EE}}}} =  \simeq {n_R^2}\int_0^\infty  {dy{S_{\ell }}\left( {N,n_R - 1/2,y n_R^2} \right)} .
\label{eq:3d_entropy_integral}
\end{equation}
Now we may expand the integrand for large $n_R$. It is also simple to show that the $n_R^2$ term of entanglement entropy receives contributions only from the integral term of formula \eqref{eq:Euler}. Therefore, the large $R$ behaviour of entanglement entropy is completely determined by equation \eqref{eq:3d_entropy_integral}.

The matrix elements of $K$ in $3+1$ dimensions are given by
\begin{align}
\frac{k_i}{\varepsilon} &= \sqrt{2 + \frac{{\ell \left( {\ell  + 1} \right) + 1/2}}{{{i^2}}} + {\mu^2}{a^2}} ,\\
{l_i} &= - \frac{{{{\left( {i + 1/2} \right)}^2}}}{{i\left( {i + 1} \right)}} \frac{1}{k_i + k_{i+1}}.
\end{align}
The integral \eqref{eq:3d_entropy_integral} can be performed explicitly. At third order in the inverse mass expansion \eqref{eq:SEEl_second_order}, we find
\begin{multline}
{S_{\mathrm{EE}}} = \left( {\frac{{3 + 2\ln \left[ {4\left( {{\mu^2}{a^2} + 2} \right)} \right]}}{{16\left( {{\mu^2}{a^2} + 2} \right)}} + \frac{{167 + 492\ln \left[ {4\left( {{\mu^2}{a^2} + 2} \right)} \right]}}{{4608{{\left( {{\mu^2}{a^2} + 2} \right)}^3}}} } \right. \\
\left. { + \frac{{-11 + 2940\ln \left[ {4\left( {{\mu^2}{a^2} + 2} \right)} \right]}}{{15360{{\left( {{\mu^2}{a^2} + 2} \right)}^5}}} + \mathcal{O}\left( {{\mu^{ - 14}}} \right)}\right){\frac{R^2}{a^2}} + \mathcal{O}\left( {{R^0}} \right) .
\label{eq:SEE_second_order_inverse_mass}
\end{multline}
The leading contribution to entanglement entropy for large entangling sphere radii is proportional to the area of the entangling sphere. This is the celebrated ``area law'' term calculated at third order in the inverse mass expansion. Using expansive techniques, we managed to acquire an analytic expression for the coefficient connecting the entangling sphere area to entanglement entropy. It is noteworthy to mention that the expansions in the inverse mass and in the size of the entangling sphere are not coinciding; the leading term in the latter expansion, i.e. the area law term, receives corrections at all orders in the inverse mass expansion.

In order to verify the validity of our expansion, we compare the perturbative results in the form of formula  \eqref{eq:SEE_second_order_inverse_mass} to the numerical calculation of entanglement entropy presented in \cite{Srednicki:1993im}, for various values of the mass parameter and $N=60$. The numerical calculation is performed with the use of Wolfram's Mathematica; the code is provided in Appendix \ref{sec:code}. It is shown in figure \ref{fig:3d_area_law} that the formula \eqref{eq:SEE_second_order_inverse_mass} is more accurate for large values of the mass parameter, as expected. Furthermore, the entanglement entropy is a decreasing function of the scalar field mass \cite{Riera:2006vj}.
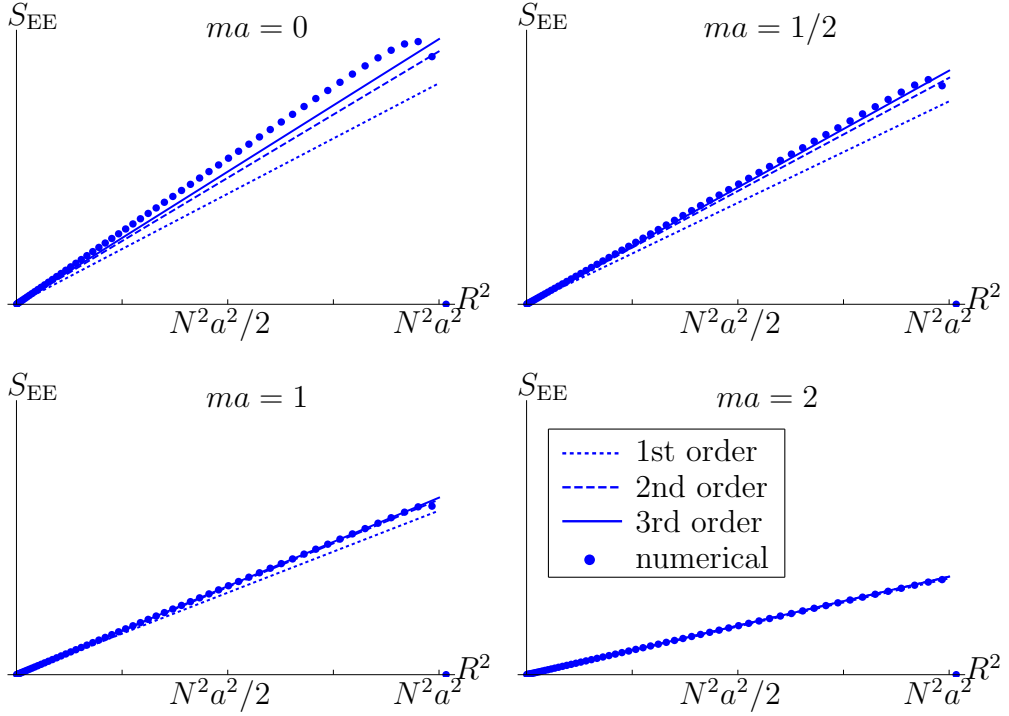
\begin{figure}[ht]
\centering
\begin{picture}(80,56)
\put(0,30){\includegraphics[width = 0.36\textwidth]{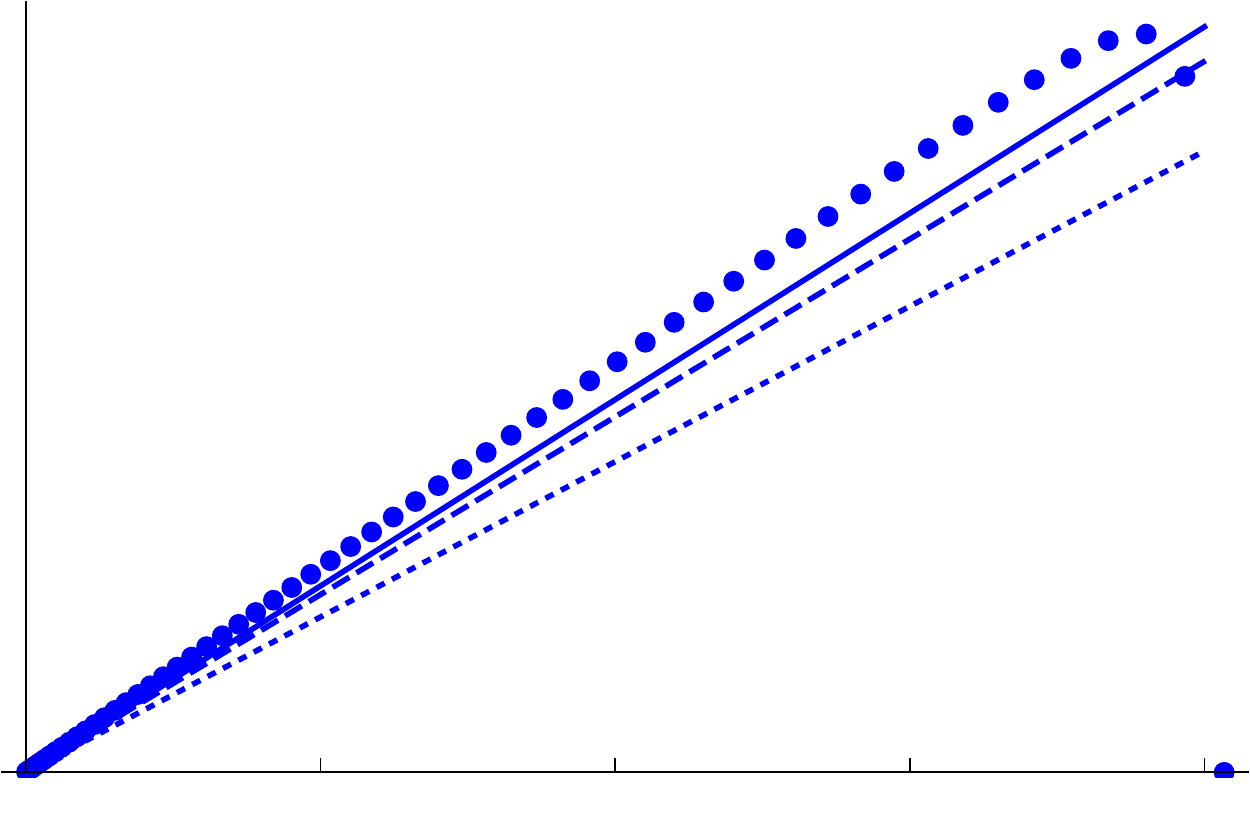}}
\put(41,30){\includegraphics[width = 0.36\textwidth]{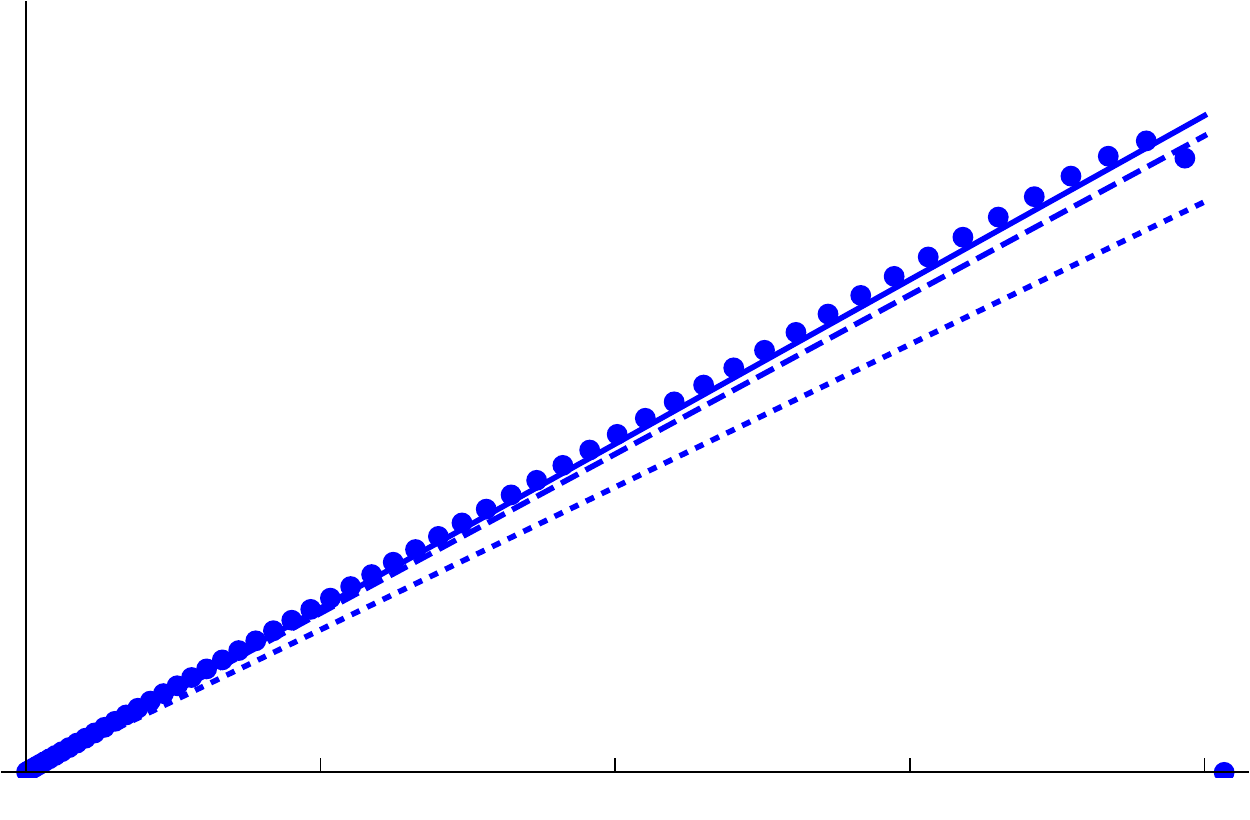}}
\put(0,0){\includegraphics[width = 0.36\textwidth]{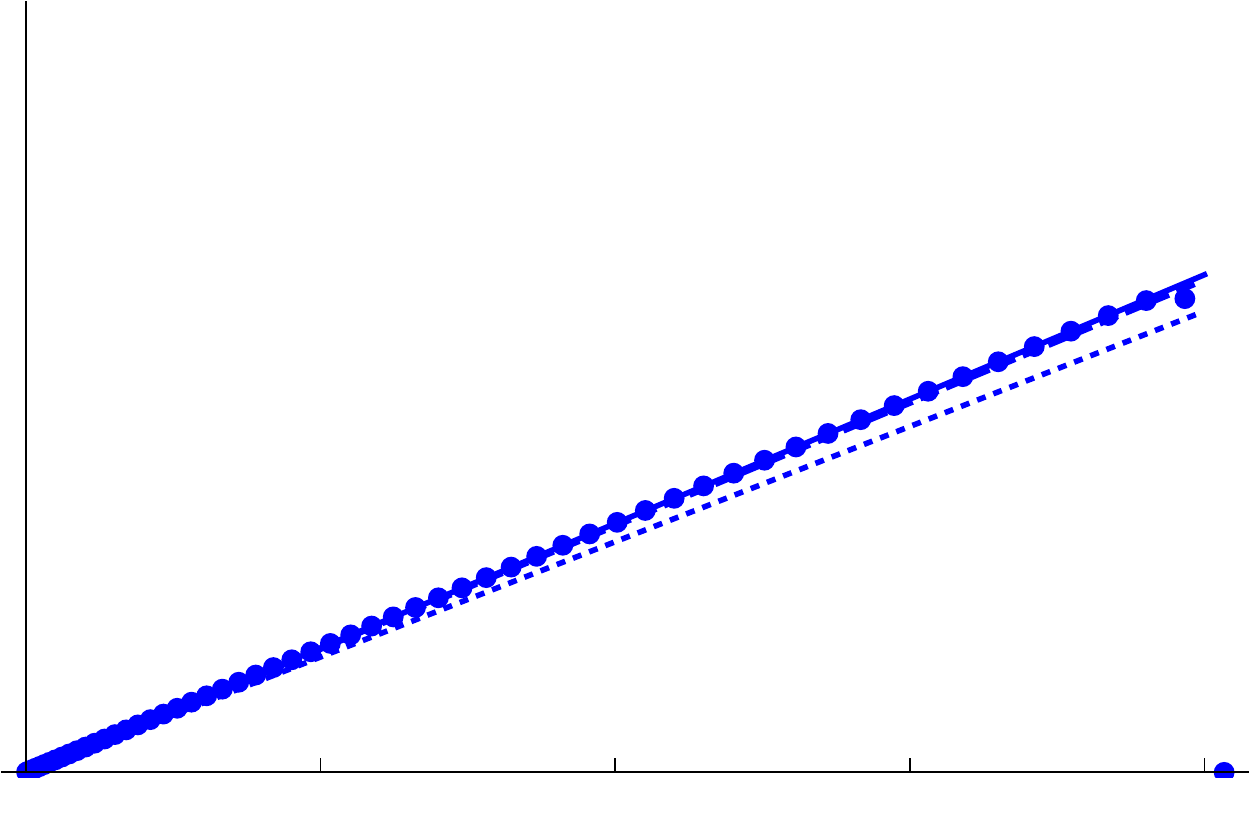}}
\put(41,0){\includegraphics[width = 0.36\textwidth]{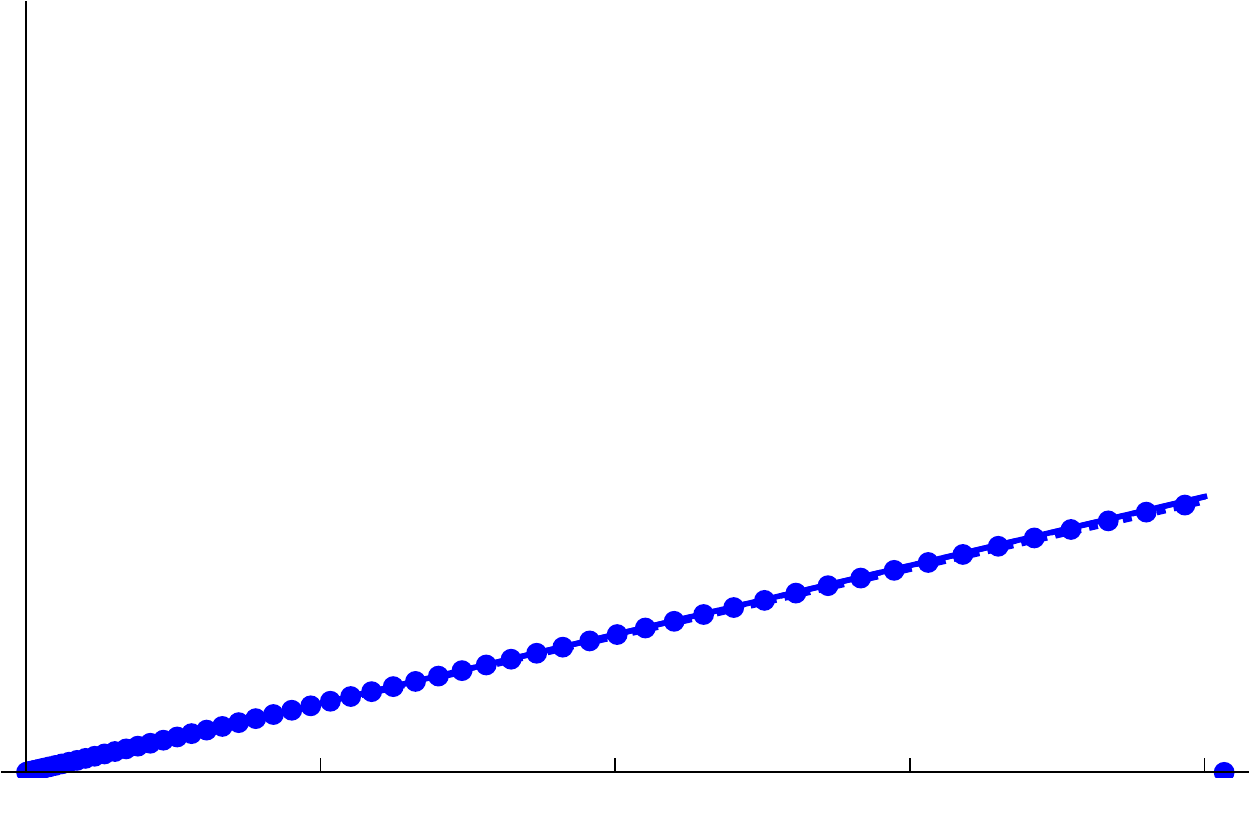}}
\put(36,31){$R^2$}
\put(77,31){$R^2$}
\put(36,1){$R^2$}
\put(77,1){$R^2$}
\put(0,54){$S_{\mathrm{EE}}$}
\put(41,54){$S_{\mathrm{EE}}$}
\put(0,24){$S_{\mathrm{EE}}$}
\put(41,24){$S_{\mathrm{EE}}$}
\put(16,53){$m a = 0$}
\put(56,53){$m a = 1/2$}
\put(16,23){$m a = 1$}
\put(57,23){$m a = 2$}
\put(13,29){$N^2 a^2 / 2$}
\put(54,29){$N^2 a^2 / 2$}
\put(13,-1){$N^2 a^2 / 2$}
\put(54,-1){$N^2 a^2 / 2$}
\put(31,29){$N^2 a^2$}
\put(72,29){$N^2 a^2$}
\put(31,-1){$N^2 a^2$}
\put(72,-1){$N^2 a^2$}
\put(43.5,9.25){\line(0,1){12}}
\put(43.5,9.25){\line(1,0){19}}
\put(62.5,9.25){\line(0,1){12}}
\put(43.5,21.25){\line(1,0){19}}
\put(44,12){\includegraphics[width = 0.07\textwidth]{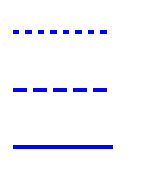}}
\put(45,8){\includegraphics[width = 0.05\textwidth]{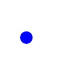}}
\put(50.5,18.625){1st order}
\put(50.5,15.75){2nd order}
\put(50.5,12.875){3rd order}
\put(50.5,10){numerical}
\end{picture}
\caption{Comparison of the numerical calculation of entanglement entropy to the perturbative formulae for the area law at first, second and third order in the inverse mass expansion. The vertical axes have the same scale for all values of the mass parameter.}
\label{fig:3d_area_law}
\end{figure}
The divergence of the numerical results from the expansive formula for entangling sphere radii close to $Na$ is an effect induced by the IR cutoff that has been imposed since the theory has been defined in a finite size spherical box. 

The numerical calculation requires the introduction of a cutoff in the values of $\ell$. The convergence of the series \eqref{eq:qft_entropy_series} gets slower as the mass parameter increases. Thus, the perturbative expansion has an additional virtue; it provides a result for entanglement entropy in cases that the numerical calculation is more difficult.

The parameter of expansion in our approach is the ration between the off-diagonal and diagonal elements of the couplings matrix $K$. This is not exactly the inverse of the mass, but rather it is equal to
\begin{equation}
\varepsilon \approx \frac{1}{\sqrt{\mu^2 a^2 + 2}}.
\end{equation}
It follows that the perturbative method can be applied even in the massless limit. Of course in such a case, the perturbation series converges much more slowly, nevertheless, it turns out that it does converge to the numerical results, as shown in figure \ref{fig:3d_area_law}.

In $3+1$ dimensions, the coefficient connecting area with entanglement entropy in massless scalar field theory has been calculated numerically in \cite{Srednicki:1993im} and improved in \cite{Lohmayer:2009sq}, found approximately equal to 0.295. Setting $\mu=0$ to the area law derived above, we find
\begin{equation}
\begin{split}
{S_{\mathrm{EE}}} &\simeq \left(\frac{{3 + 2\ln 8}}{{32}}+ \frac{{167 + 492\ln 8}}{{36864}} + \frac{{-11 + 2940\ln 8}}{{491520}} \right)  {\frac{R^2}{a^2}} \\ &\simeq \left( 0.224 + 0.032 + 0.012 \right) {\frac{R^2}{a^2}}  \simeq 0.268{\frac{R^2}{a^2}} ,
\end{split}
\end{equation}
which is a good approximation to the numerical result and can be further improved continuing the perturbative expansion at higher orders.

\subsubsection*{$2+1$ Dimensions}
In $2+1$ dimensions, the entanglement entropy equals the sum of all $\ell$ sectors, as shown by equation \eqref{eq:qft_entropy_series_d2}. With the use of Euler-Maclaurin formula \eqref{eq:Euler}, it may be approximated by the integral
\begin{equation}
{S_{{\rm{EE}}}} = \sum\limits_{\ell  = -\infty}^\infty  {{S_{\ell }}\left(N, n, \ell^2 \right)}  \simeq \int_{-\infty}^\infty  {d\ell {S_{\ell }}\left(N, n,\ell^2 \right)} .
\end{equation}

\begin{figure}[ht]
\centering
\begin{picture}(80,56)
\put(0,30){\includegraphics[width = 0.36\textwidth]{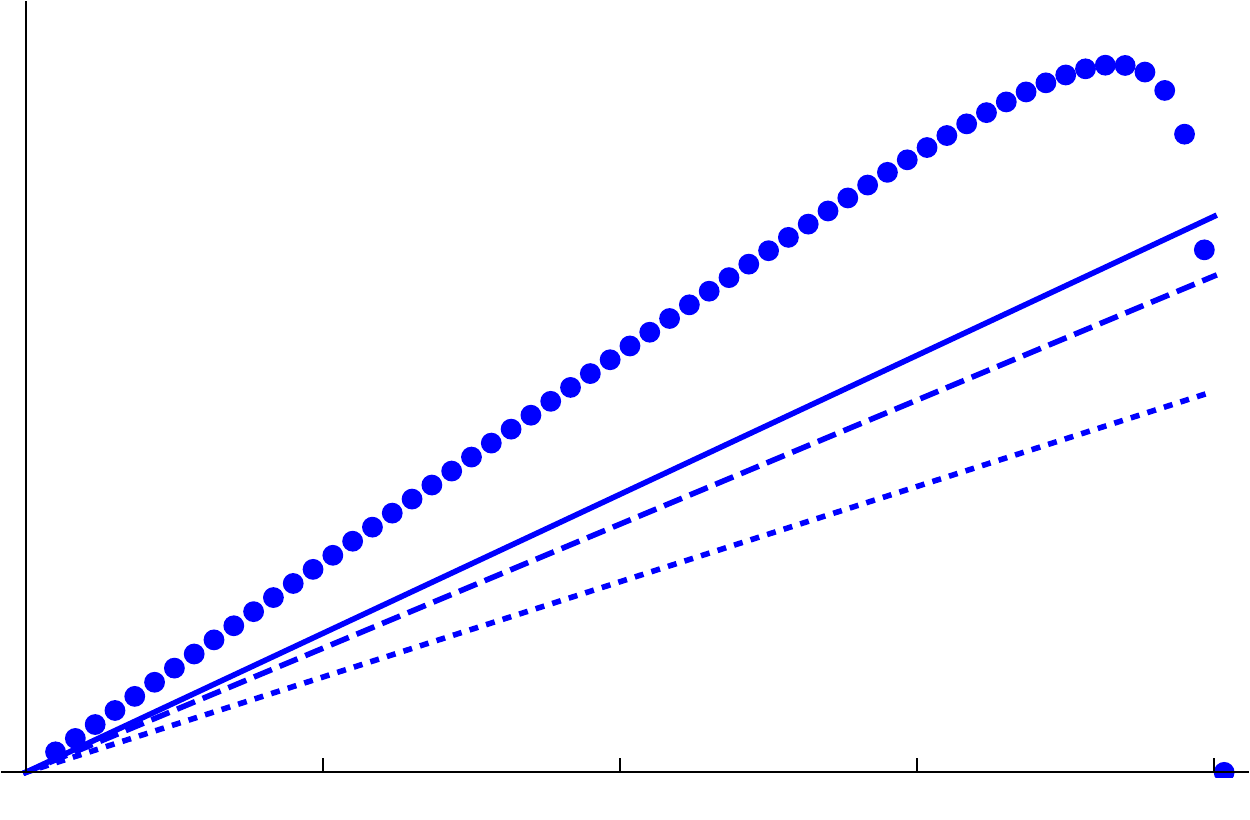}}
\put(41,30){\includegraphics[width = 0.36\textwidth]{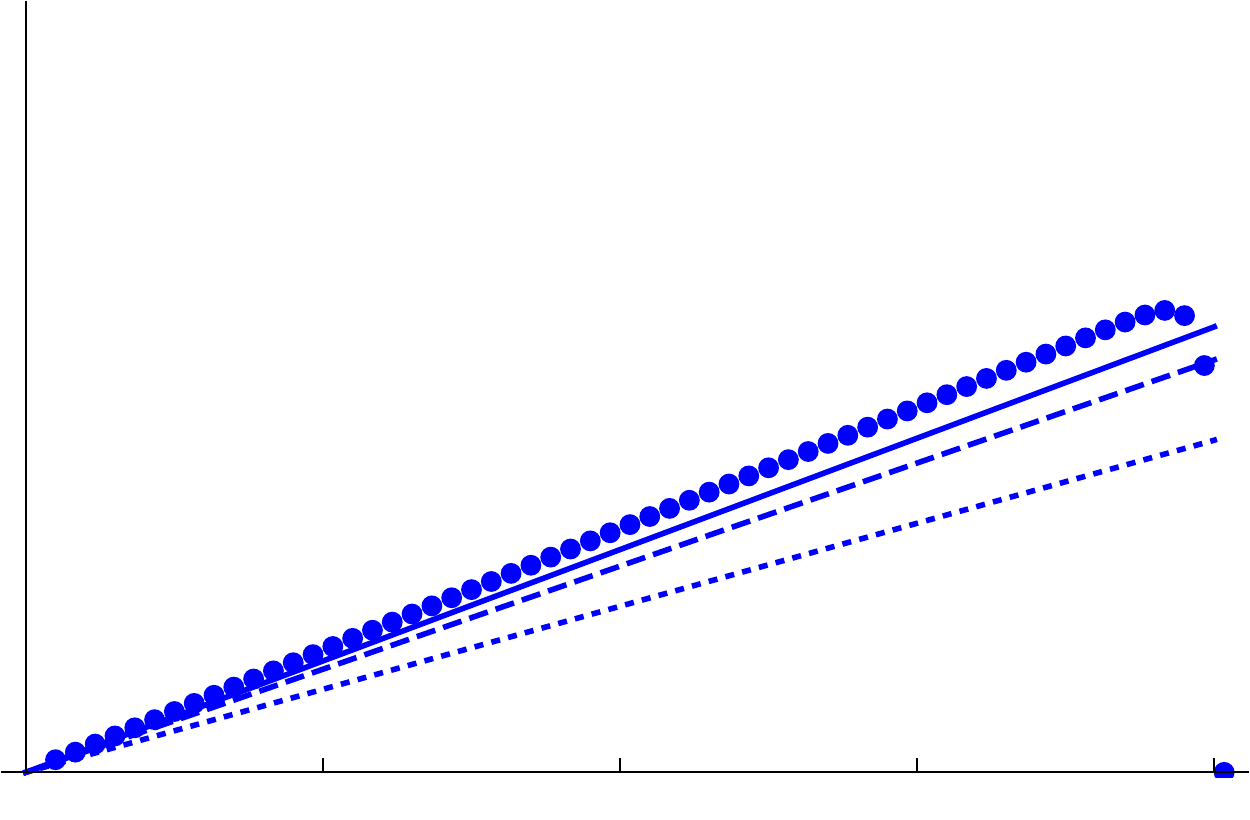}}
\put(0,0){\includegraphics[width = 0.36\textwidth]{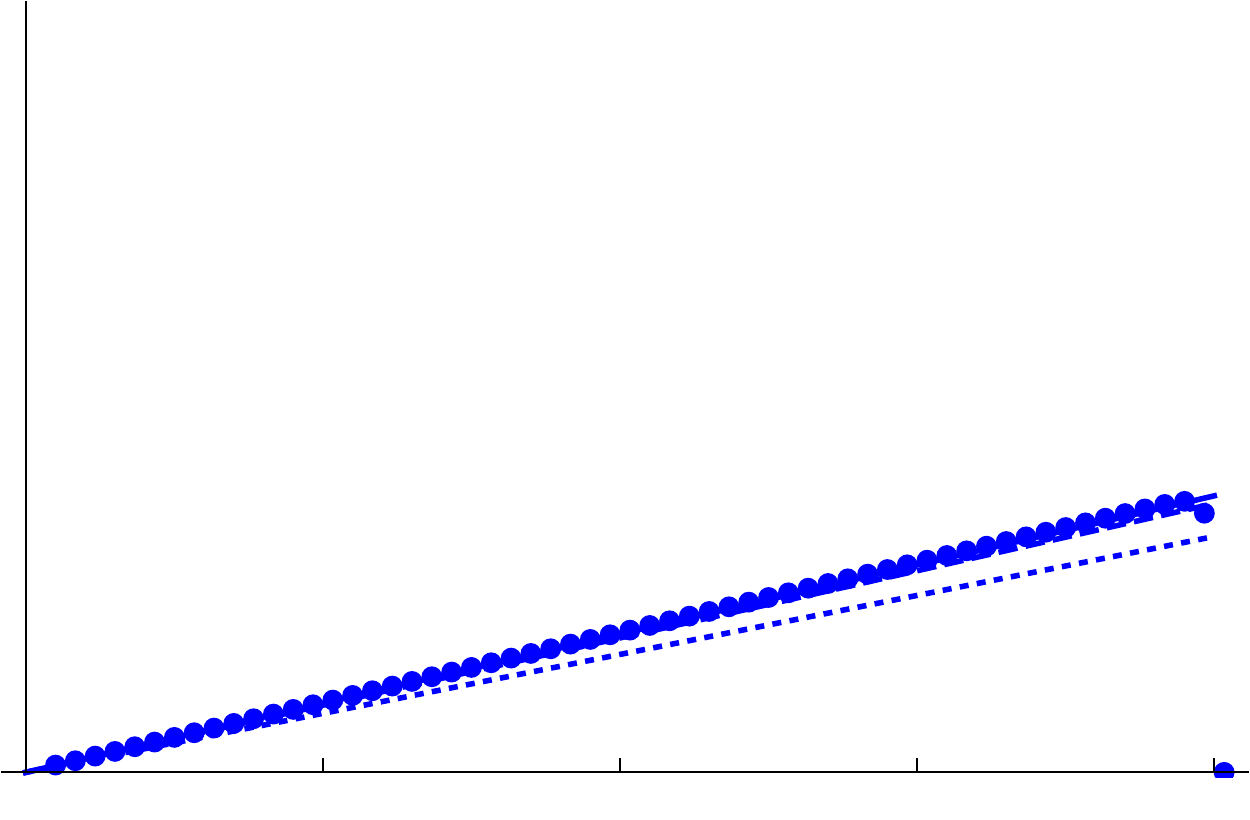}}
\put(41,0){\includegraphics[width = 0.36\textwidth]{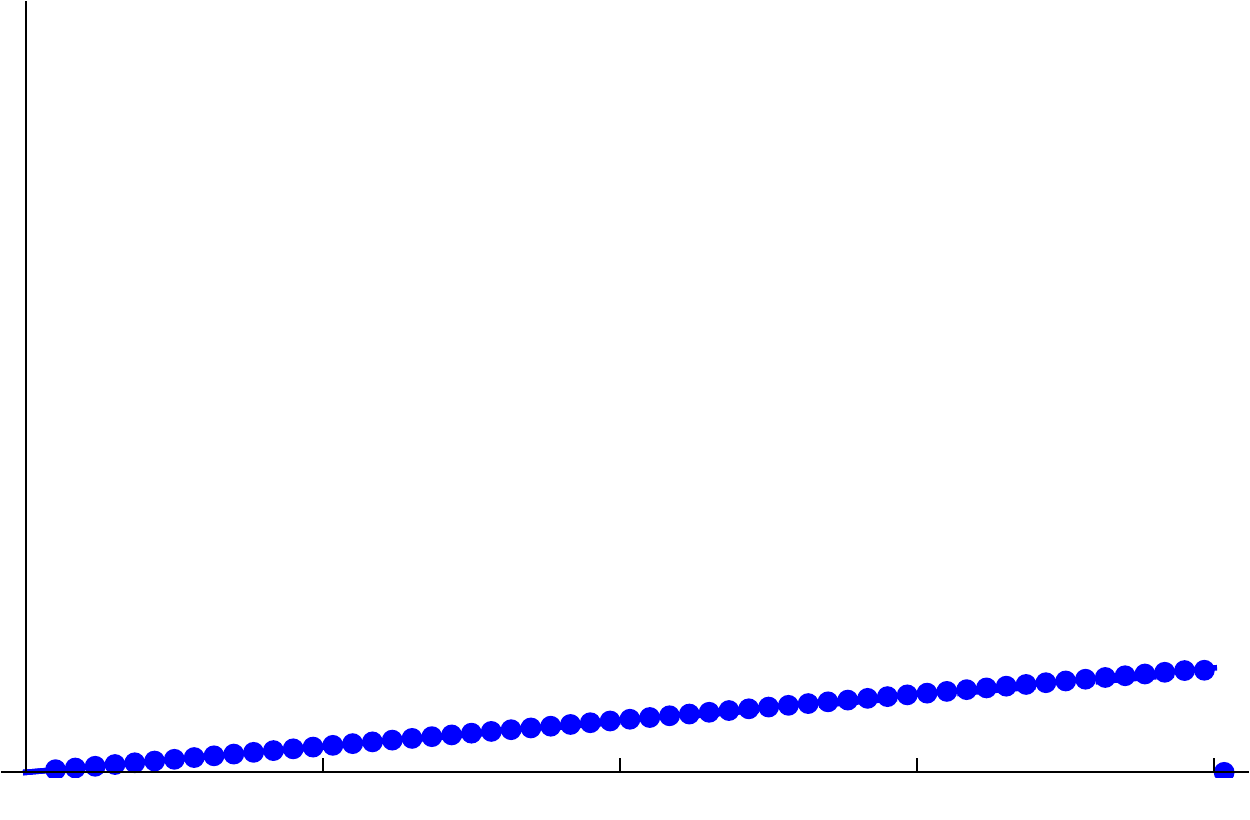}}
\put(36,31){$R$}
\put(77,31){$R$}
\put(36,1){$R$}
\put(77,1){$R$}
\put(0,54){$S_{\mathrm{EE}}$}
\put(41,54){$S_{\mathrm{EE}}$}
\put(0,24){$S_{\mathrm{EE}}$}
\put(41,24){$S_{\mathrm{EE}}$}
\put(16,53){$m a = 0$}
\put(56,53){$m a = 1/2$}
\put(16,23){$m a = 1$}
\put(57,23){$m a = 2$}
\put(14,29){$N a / 2$}
\put(55,29){$N a / 2$}
\put(14,-1){$N a / 2$}
\put(55,-1){$N a / 2$}
\put(33,29){$N a$}
\put(74,29){$N a$}
\put(33,-1){$N a$}
\put(74,-1){$N a$}
\put(43.5,9.25){\line(0,1){12}}
\put(43.5,9.25){\line(1,0){19}}
\put(62.5,9.25){\line(0,1){12}}
\put(43.5,21.25){\line(1,0){19}}
\put(44,12){\includegraphics[width = 0.07\textwidth]{3dlegend.pdf}}
\put(45,8){\includegraphics[width = 0.05\textwidth]{3dlegenddot.pdf}}
\put(50.5,18.625){1st order}
\put(50.5,15.75){2nd order}
\put(50.5,12.875){3rd order}
\put(50.5,10){numerical}
\end{picture}
\caption{Comparison of the numerical calculation of entanglement entropy to the perturbative area law formulae at first, second and third order in the inverse mass expansion. The vertical axes have the same scale for all values of the mass parameter.}
\label{fig:2d_area_law}
\end{figure}

As in $3+1$ dimensions, in order to find the asymptotic behaviour of this integral for large entangling circles, we perform the change of variables $\ell = y n_R$,
\begin{equation}
{S_{{\rm{EE}}}} =  \simeq {n_R}\int_{-\infty}^\infty  {dy{S_{\ell }}\left( {N,n_R-1/2,n_R^2 y^2} \right)} .
\end{equation}
Now we may expand the integrand for large $n_R$. In $2+1$ dimensions, the matrix elements of $K$ are given by
\begin{align}
{k_i} &= \sqrt{2 + \frac{{\ell^2}}{{{i^2}}} + {\mu^2}{a^2}} ,\\
{l_i} &= - \frac{{ {i + 1/2} }}{{\sqrt{i\left( {i + 1} \right)}}} \frac{1}{k_i + k_{i+1}}.
\end{align}
At third order in the inverse mass expansion, using formula \eqref{eq:SEEl_second_order} we obtain
\begin{multline}
S_{\rm{EE}} = \left(\frac{{ {-1 + 2\ln \left[ {16\left( {{\mu^2}{a^2} + 2} \right)} \right]} }}{{32\left( {{\mu^2}{a^2} + 2} \right)^{3/2}}} +\frac{{ {-3019 + 2460\ln \left[ {16\left( {{\mu^2}{a^2} + 2} \right)} \right]} }}{{24576\left( {{\mu^2}{a^2} + 2} \right)^{7/2}}} \right.\\
\left. +7 \frac{{ {-6593 + 4410\ln \left[ {16\left( {{\mu^2}{a^2} + 2} \right)} \right]} }}{{131072\left( {{\mu^2}{a^2} + 2} \right)^{11/2}}} + \mathcal{O} \left( \mu^{-15} \right) \right){\frac{\pi R}{a}} + \mathcal{O}\left( {{R^{-1}}} \right).
\label{eq:SEE_area_law_2nd_order_2d}
\end{multline}
Figure \ref{fig:2d_area_law} compares the perturbative formula \eqref{eq:SEE_area_law_2nd_order_2d} to the numerical calculation of entanglement entropy with $N=60$ for various values of the mass parameter. As expected, the perturbative results are more accurate for larger values of the mass parameter. In general the perturbative series converges more slowly than in $3+1$ dimensions.

In the massless case, we yield
\begin{equation}
\begin{split}
{S_{{\rm{EE}}}} &\simeq \left( {\frac{{ - 1 + 2\ln 32}}{{64\sqrt 2 }} + \frac{{ - 3019 + 2460\ln 32}}{{196608\sqrt 2 }} + 7\frac{{ - 6593 + 4410\ln 32}}{{4194304\sqrt 2 }}} \right)\frac{{\pi R}}{a}\\
 &\simeq \left( {0.206 + 0.062 + 0.032} \right)\frac{R}{a} \simeq 0.300\frac{R}{a} .
\end{split}
\end{equation}
As in $3+1$ dimensions, the perturbation series converges to the numerical results even in the massless case.

\subsubsection*{$1+1$ Dimensions}
\begin{figure}[ht]
\centering
\begin{picture}(80,56)
\put(0,30){\includegraphics[width = 0.36\textwidth]{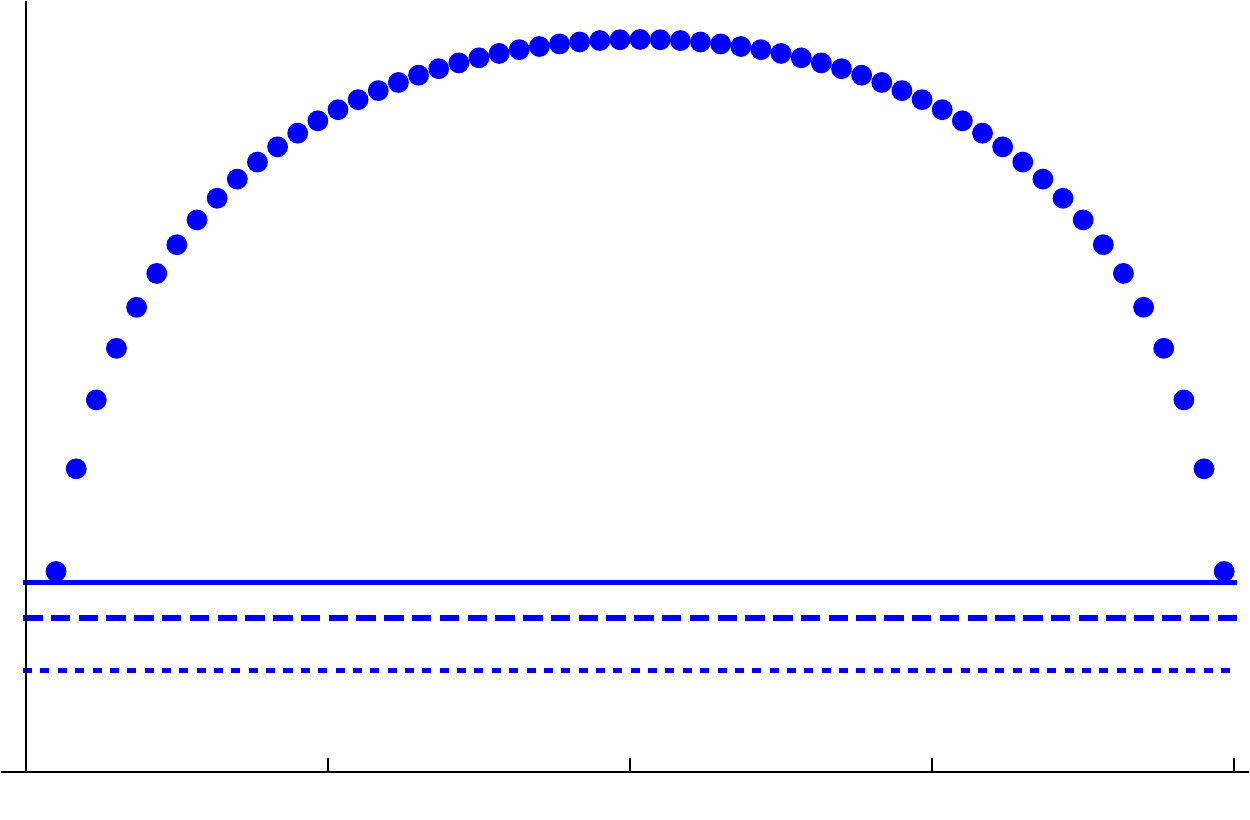}}
\put(41,30){\includegraphics[width = 0.36\textwidth]{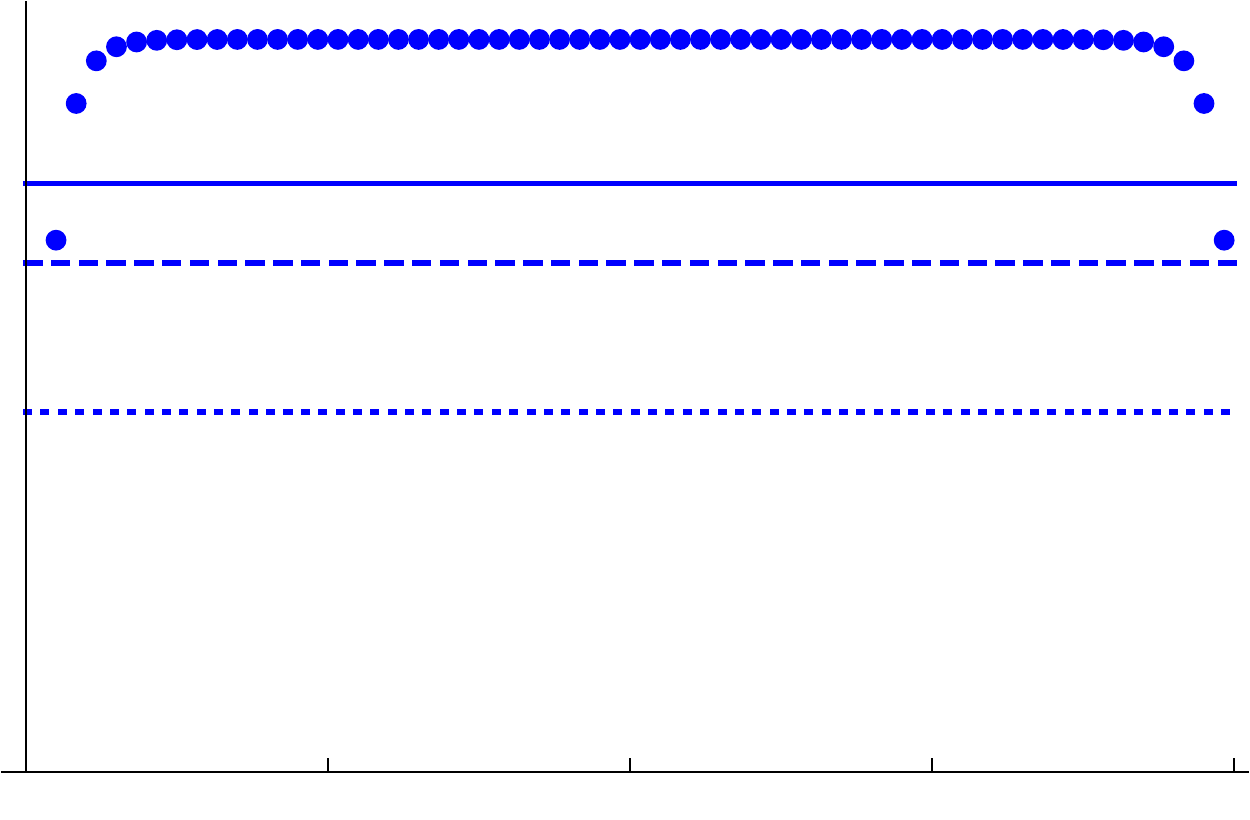}}
\put(0,0){\includegraphics[width = 0.36\textwidth]{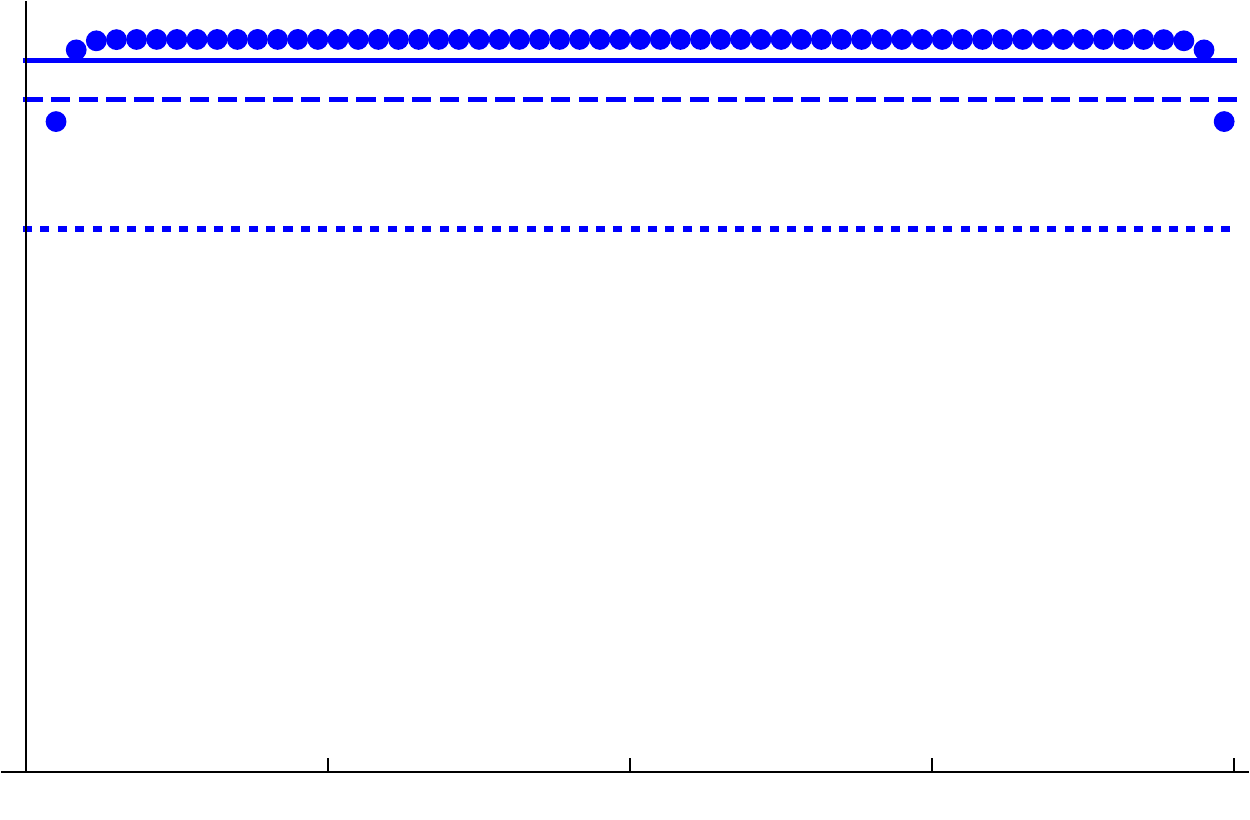}}
\put(41,0){\includegraphics[width = 0.36\textwidth]{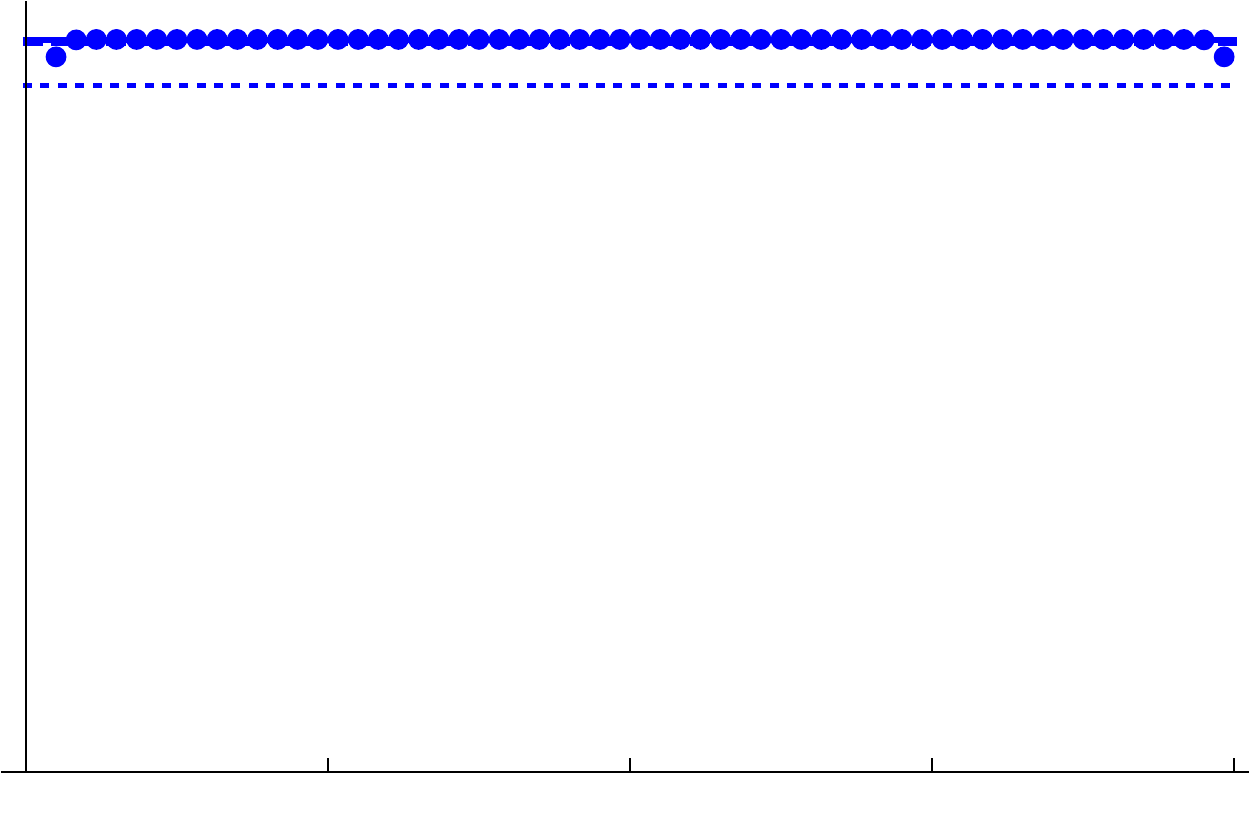}}
\put(36,31){$R$}
\put(77,31){$R$}
\put(36,1){$R$}
\put(77,1){$R$}
\put(0,54){$S_{\mathrm{EE}}$}
\put(41,54){$S_{\mathrm{EE}}$}
\put(0,24){$S_{\mathrm{EE}}$}
\put(41,24){$S_{\mathrm{EE}}$}
\put(16,53){$m a = 0$}
\put(56,53){$m a = 1/2$}
\put(16,23){$m a = 1$}
\put(57,23){$m a = 2$}
\put(15,29){$N a / 2$}
\put(56,29){$N a / 2$}
\put(15,-1){$N a / 2$}
\put(56,-1){$N a / 2$}
\put(33,29){$N a$}
\put(74,29){$N a$}
\put(33,-1){$N a$}
\put(74,-1){$N a$}
\put(43.5,6.25){\line(0,1){12}}
\put(43.5,6.25){\line(1,0){19}}
\put(62.5,6.25){\line(0,1){12}}
\put(43.5,18.25){\line(1,0){19}}
\put(44,9){\includegraphics[width = 0.07\textwidth]{3dlegend.pdf}}
\put(45,5){\includegraphics[width = 0.05\textwidth]{3dlegenddot.pdf}}
\put(50.5,15.625){1st order}
\put(50.5,12.75){2nd order}
\put(50.5,9.875){3rd order}
\put(50.5,7){numerical}
\end{picture}
\caption{Comparison of the numerical results for entanglement entropy to the first, second and third order inverse mass expansion. The vertical axes do not have the same scale, entanglement entropy is a decreasing function of the mass parameter, as in higher dimensions.}
\label{fig:1d_area_law}
\end{figure}

In $1+1$ dimensions, the matrix elements of $K$ are given by
\begin{align}
{k_i} &= \sqrt{2 + {\mu^2}{a^2}} ,\\
{l_i} &= - \frac{1}{k_i + k_{i+1}}.
\end{align}
At third order in the inverse mass expansion, we obtain
\begin{multline}
{S_{{\rm{EE}} }} = \left( {\frac{{1 + 2 \ln \left[ {4{{\left( {2 + {\mu^2 a^2} } \right)}}} \right]}}{{16{{\left( {2 + {\mu^2 a^2} } \right)}^2}}} + \frac{{1 + 164\ln \left[ {4{{\left( {2 + {\mu^2 a^2} } \right)}}} \right]}}{{512{{\left( {2 + {\mu^2 a^2} } \right)}^4}}} } \right. \\
\left. {+ \frac{{-599 + 2940\ln \left[ {4{{\left( {2 + {\mu^2 a^2} } \right)}}} \right]}}{{3072{{\left( {2 + {\mu^2 a^2} } \right)}^6}}} + \mathcal{O}\left( {{\mu^{ - 16}}} \right)}\right){n_R^0} + \mathcal{O}\left( {{n_R^{ - 2}}} \right) .
\end{multline}
In figure \ref{fig:1d_area_law}, the comparison of the perturbative formulae to the numerical calculation of the entanglement entropy is depicted. The series converges more slowly than in higher dimensions.

Especially in the massless case, the perturbative formulae fail completely to capture the logarithmic behaviour of entanglement entropy (figure \ref{fig:1d_area_law} top-left). Technically, this happens due to the structure of the couplings matrix $K$. In all cases this matrix is diagonally dominant, i.e. the sum of the absolute value of all non-diagonal elements does not exceed the diagonal one, in all rows and columns. As a result, the perturbative calculation of its square root and its inverse converges. Only in $1+1$ dimensions and only in the massless case, the matrix saturates the diagonally dominant criterion. Not unexpectedly, the saturating case, lying between convergence and divergence, leads to a logarithmic dependence on the cutoff scale. However, this logarithmic dependence cannot be evident in a finite number of terms of the perturbation series. We will return to this kind of behaviour in the section \ref{subsec:EE_subleading} on the subleading contributions to entanglement entropy.

The area law is the leading contribution to the entanglement entropy for large entangling sphere radii in all number of dimensions. The reason for this fact can be attributed to the locality of the underlying field theory \cite{Plenio:2004he,Cramer:2005mx,Eisert:2008ur}. The locality is depicted to the fact that the matrix $K$ contains interaction elements only in the subdiagonal and superdiagonal. As a result, no matter what is the size of the entangling sphere (the value of $n$), there is only one element of $K$ connecting a degree of freedom of subsystem $A$ to a degree of freedom of subsystem $A^C$. This property is inherited to the leading corrections in matrix $B$, and, thus, to the eigenvalues of the reduced density matrix. Had the theory been non-local, the number of leading contributions to entanglement entropy, would be a complicated function of the entangling sphere radius in general, leading to large divergences from the area law. In a more geometric phrasing, the area law emerges from locality, since the pairs of strongly correlated degrees of freedom (i.e. neighbours) that have been separated by the entangling surface, are proportional to its area.

\subsection{Beyond the Area Law}
\label{subsec:EE_subleading}

The area law term of entanglement entropy is the leading contribution to the entanglement entropy for large radii of the entangling sphere. There are also subleading terms, which can also be calculated in the inverse mass expansion that we developed in section \ref{sec:expansion}.

In $2+1$ and $1+1$ dimensions, the subleading terms vanish as $a \to 0$. We will not extend our analysis to these cases. In the case of $3+1$ dimensions, the first subleading term is a constant. There are two contributions to this term. The first one comes from the integral term \eqref{eq:3d_entropy_integral} and can be acquired by appropriate expansion of the integrand. The second contribution comes from the rest of the terms in Euler-Maclaurin formula \eqref{eq:Euler}. Taking into account that $\mathop {\lim }\limits_{\ell  \to \infty } \left( {2\ell  + 1} \right){S_{{\rm{EE}}\ell }} = 0$, the equation \eqref{eq:Euler} reads
\begin{equation}
{S_{{\rm{EE}}}} = \int_0^\infty  {d\ell \left( {2\ell  + 1} \right){S_{\ell }}}  + \frac{{{S_{0}}}}{2} - \sum\limits_{k = 1}^\infty  {\frac{{{B_{2k}}}}{{\left( {2k} \right)!}}{{\left. {\frac{{{d^{2k - 1}}\left( {2\ell  + 1} \right){S_{\ell }}}}{{d{\ell ^{2k - 1}}}}} \right|}_{\ell  = 0}}} .
\end{equation}
Since the parameter $\ell$ appears in ${S_{\ell }}$ only in the form of the fraction $\ell \left( \ell + 1 \right) / n_R^2$, any action of the derivative operator on ${S_{\ell }}$ results in a term two orders smaller in the $n_R$ expansion. This implies that apart from the ${{{S_{0}}}}/{2}$ term, we have only one more contribution at $n_R^0$ order, namely the $k = 1$ term, and specifically the part of latter where the derivative acts on the factor $2\ell + 1$ and not on ${S_{\ell }}$. Bearing in mind that $B_2 = 1 / 6$, the contribution to the constant term by the discrete part of the Euler-Maclaurin formula is ${{{S_{0}}}}/{3}$.

Performing the expansion of the integrand of equation \eqref{eq:3d_entropy_integral} using the ${S_{\ell }}$ acquired at second order in the inverse mass expansion \eqref{eq:SEEl_second_order} and taking into account the extra ${S_{0 } / 3}$ contribution to the constant term, we find
\begin{multline}
{S_{{\rm{EE}}}} = S_{{\rm{EE}}}^{\left( 2 \right)}\frac{{{R^2}}}{{{a^2}}} - \left( \frac{1}{{48\left( {2 + {\mu^2}{a^2}} \right)}} + \frac{{1 + 2\ln \left( {4\left( {2 + {\mu^2}{a^2}} \right)} \right)}}{{96{{\left( {2 + {\mu^2}{a^2}} \right)}^2}}} \right.\\
\left. + \frac{{127 - 90\ln \left( {4\left( {2 + {\mu^2}{a^2}} \right)} \right)}}{{9600{{\left( {2 + {\mu^2}{a^2}} \right)}^3}}} + \frac{{1 + 164\ln \left( {4\left( {2 + {\mu^2}{a^2}} \right)} \right)}}{{3072{{\left( {2 + {\mu^2}{a^2}} \right)}^4}}} + \mathcal{O} \left( \mu^{10} \right)\right) + \mathcal{O}\left( {{R^{ - 2}}} \right) .
\end{multline}

\begin{figure}[ht]
\centering
\begin{picture}(60,35)
\put(5,0){\includegraphics[width = 0.52\textwidth]{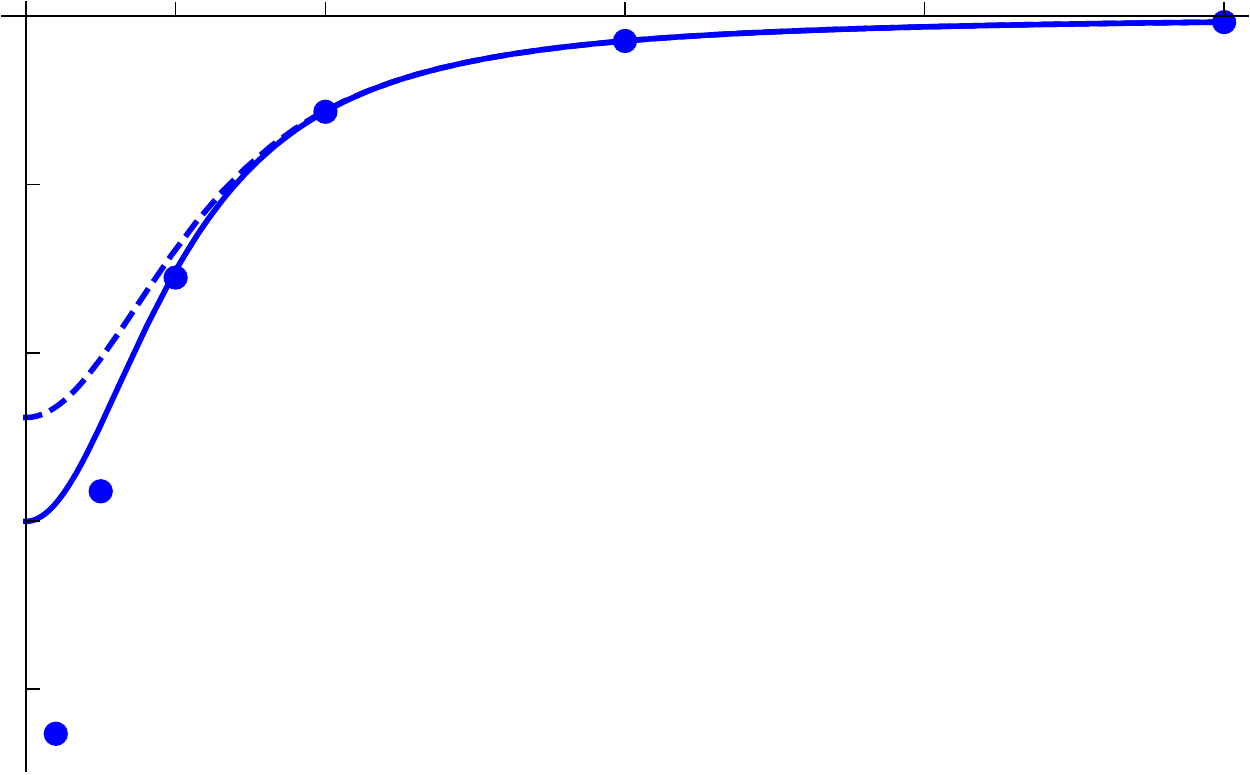}}
\put(57.25,31){$\mu^2 a^2$}
\put(4.5,33.25){$S_{\mathrm{EE}}^{\left( 0 \right)}$}
\put(0,2.75){-0.04}
\put(0,17){-0.02}
\put(17.75,32.5){2}
\put(30.25,32.5){4}
\put(42.75,32.5){6}
\put(55.25,32.5){8}
\put(28.5,7.25){\line(0,1){10}}
\put(28.5,7.25){\line(1,0){19}}
\put(47.5,7.25){\line(0,1){10}}
\put(28.5,17.25){\line(1,0){19}}
\put(29,10){\includegraphics[width = 0.07\textwidth]{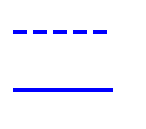}}
\put(30,6.25){\includegraphics[width = 0.05\textwidth]{3dlegenddot.pdf}}
\put(35.5,14.125){1st order}
\put(35.5,11.125){2nd order}
\put(35.5,8.25){numerical}
\end{picture}
\caption{The subleading constant term of entanglement entropy in scalar field theory in $3+1$ dimensions, as function of the mass parameter}
\label{fig:subleading}
\end{figure}

In order to compare the constant subleading term found above to the numerical calculation of entanglement entropy, we performed a linear fit to the outcome of the numerical calculation of the form $S_{\mathrm{EE}} = c_2 n_R^2 + c_0$, for various values of the parameter $\mu^2 a^2$. The perturbative formulae indeed approximate the numerical results well, as shown in figure \ref{fig:subleading}, apart from the massless limit.

At finite order in the inverse mass expansion, the first subleading term is a constant, even in the massless case. The usual treatment of entanglement entropy in $3+1$ dimensions in either conformal field theory or in theories with holographic duals through the Ryu-Takayanagi conjecture predicts an expansion for entanglement entropy of the form
\begin{equation}
{S_{{\rm{EE}}}} = {c_2}\frac{{{R^2}}}{{{a^2}}} + {c_0} + {c} \ln \frac{a}{R} + \mathcal{O}\left( {{a^{ - 2}}} \right) .
\label{eq:energy_cutoff_pattern}
\end{equation}
So, how is the absence of the logarithmic term in our expansion explained? In the case of massive scalar field theory, the answer is the existence of a fundamental scale in the theory, that of the mass, which naturally cutoffs the logarithmic term. As far as the massless limit in $3+1$ dimensions is concerned, the reason is more complicated and related to the failure to capture the leading entanglement entropy contribution in the same limit in $1+1$ dimensions. In a similar manner our perturbation theory is unable to capture the constant term in massless $2+1$ field theory. From a technical point of view, we understand this failure of our perturbation theory as follows:

The formulae used in our perturbation theory for the square root of matrix $K$, as well as the formulae for the inverse of matrices $A$ and $C$, present some ``edge effects'' due to the fact that the matrices used in the inverse mass expansion are of finite size. This can be seen in the form of the factors $1-\delta_{i1}$ and $1-\delta_{i,N-1}$ in formula \eqref{eq:edge_effect_omega} or the factor $1-\delta_{in}$ in formula \eqref{eq:edge_effect_2}. Such ``edge effects'' can be treated analytically for arbitrary $N$ and $n$ in our expansion, as long as the order of the expansion is kept lower than the dimension of the matrices. If this is not the case, these ``edge effects'' will propagate through the matrix and eventually will get reflected at the opposite ends of the matrices that generate them, resulting in spreading all over the matrix elements. This qualitative behaviour implies the following:

\begin{itemize}
\item The reflections of these ``edge effects'' will lead to matrices $\Omega$, $A^{-1}$, etc, that depend on all the elements of the matrix $K$. Therefore, at high orders in the perturbation theory, such reflections introduce contributions to the entanglement entropy that depend on the global characteristics of the entangling surface. Such ``universal'' terms cannot be captured at any finite order in our perturbation series. They are rather non-perturbative effects in this expansion. The logarithmic term in even number of spacetime dimensions \cite{Ryu:2006bv,Ryu:2006ef,Casini:2011kv,Myers:2010xs,Myers:2010tj,Solodukhin:2008dh,Casini:2010kt}, as well as the constant term in odd number of dimensions \cite{Myers:2010xs,Myers:2010tj} are known to be exactly this kind of universal terms, and, thus, our inability to capture them in the inverse mass expansion should not be considered surprising. Of course such effects are visible in the numerical calculations.
\item The terms we capture in our perturbation series cannot sense the global properties of the region defined by the entangling surface. They have the property to depend on the local characteristics of the entangling surface. In a more technical language, this is depicted to the fact that the perturbative expressions for the elements of the matrices $\Omega$, $A^{-1}$ and $C^{-1}$ depend on a finite number of the elements of matrix $K$. This is the reason our method is appropriate to capture the ``area law'', as well as subleading terms that scale with smaller powers of the entangling sphere radius. Therefore, our method is appropriate to study the dependence of such terms on geometric characteristics of the entangling surface, such as curvature \cite{Solodukhin:2008dh}, if more general entangling surfaces are considered.
\item The introduction of a mass exponentially dumps the propagation of the ``edge effects'' through the matrix elements \cite{Brandao:2014ppa}. As a result, our expansive calculations accurately converge to the numerical calculations in this case.
\end{itemize}

\subsection{Dependence on the Regularization Scheme}
\label{subsec:regularization}
Finally, we would like to comment on the dependence of the area law term, as well as the subleading terms of entanglement entropy on the regularization scheme. In our analysis, we have applied a peculiar, inhomogeneous regularization. Namely, we have imposed a cutoff in the radial direction, but not in the angular directions. Thus, the measurables that we have calculated, are those measured by a peculiar observer who has access to radial excitations of the theory up to an energy scale $1/a$ and to arbitrary high energy azimuthal excitations.

We could have applied a different more homogeneous regularization imposing an azimuthal cutoff by constraining the summation series in $\ell$ to a maximum value equal to $\ell_{\max}$. Such a prescription would make our approach more similar to a traditional square lattice regularization. Notice however, that even in the square lattice case, the imposed cutoff is a cutoff to each of the momentum components and not strictly an energy cutoff that would allow direct comparison with formulae like \eqref{eq:energy_cutoff_pattern}.

As we discussed above, locality enforces the area law term to depend on the characteristics of the underlying theory in the region of the entangling surface. Therefore, a natural selection for an azimuthal cutoff $\ell_{\max}$, when considering a $d$-dimensional entangling surface should have the following property: the total number of harmonics with $\ell \le \ell_{\max}$ should equal the area of the entangling surface divided by $a^d$. In $3+1$ dimensions this argument implies that a natural selection for the azimuthal cutoff is $\ell_{\max} = 2 \sqrt{\pi} R / a$, whereas in $2+1$ dimensions it implies $\ell_{\max} = \pi R / a$. In all number of dimensions such a cutoff is of the form $\ell_{\max} = c R / a$, where $c$ is a constant. It is not difficult to repeat our analysis including this azimuthal cutoff. The only extra necessary steps are the introduction of a finite upper bound in the definite integral \eqref{eq:3d_entropy_integral} and similarly the inclusion of the terms calculated at $x=\ell_{\max}$ in the Euler-Maclaurin formula \eqref{eq:Euler}.

As an indicative example, in $3+1$ dimensions, the area law term calculated at second order in the inverse mass expansion assumes the form
\begin{multline}
{S_{{\rm{EE}}}} = \left( {\frac{{3 + 2\ln \left[ {4\left( {{\mu ^2}{a^2} + 2} \right)} \right]}}{{\left( {{\mu ^2}{a^2} + 2} \right)}} - \frac{{3 + 2\ln \left[ {4\left( {{\mu ^2}{a^2} + 2 + {c^2}} \right)} \right]}}{{\left( {{\mu ^2}{a^2} + 2 + {c^2}} \right)}}} \right.\\
\left. {\frac{{167 + 492\ln \left[ {4\left( {{\mu ^2}{a^2} + 2} \right)} \right]}}{{4608{{\left( {{\mu ^2}{a^2} + 2} \right)}^3}}} - \frac{{167 + 492\ln \left[ {4\left( {{\mu ^2}{a^2} + 2 + {c^2}} \right)} \right]}}{{4608{{\left( {{\mu ^2}{a^2} + 2 + {c^2}} \right)}^3}}} + \mathcal{O}\left( {{\mu ^{ - 10}}} \right)} \right)\frac{{{R^2}}}{{{a^2}}} .
\label{eq:SEE_c}
\end{multline}
This equation implies that the coefficient of the area law term depends on the regularization scheme. The coefficients calculated in section \ref{subsec:area_law}, which correspond to the selection $c \to \infty$, serve as an upper bound for the area law coefficient.

\begin{figure}[ht]
\centering
\begin{picture}(80,35)
\put(5,0){\includegraphics[width = 0.52\textwidth]{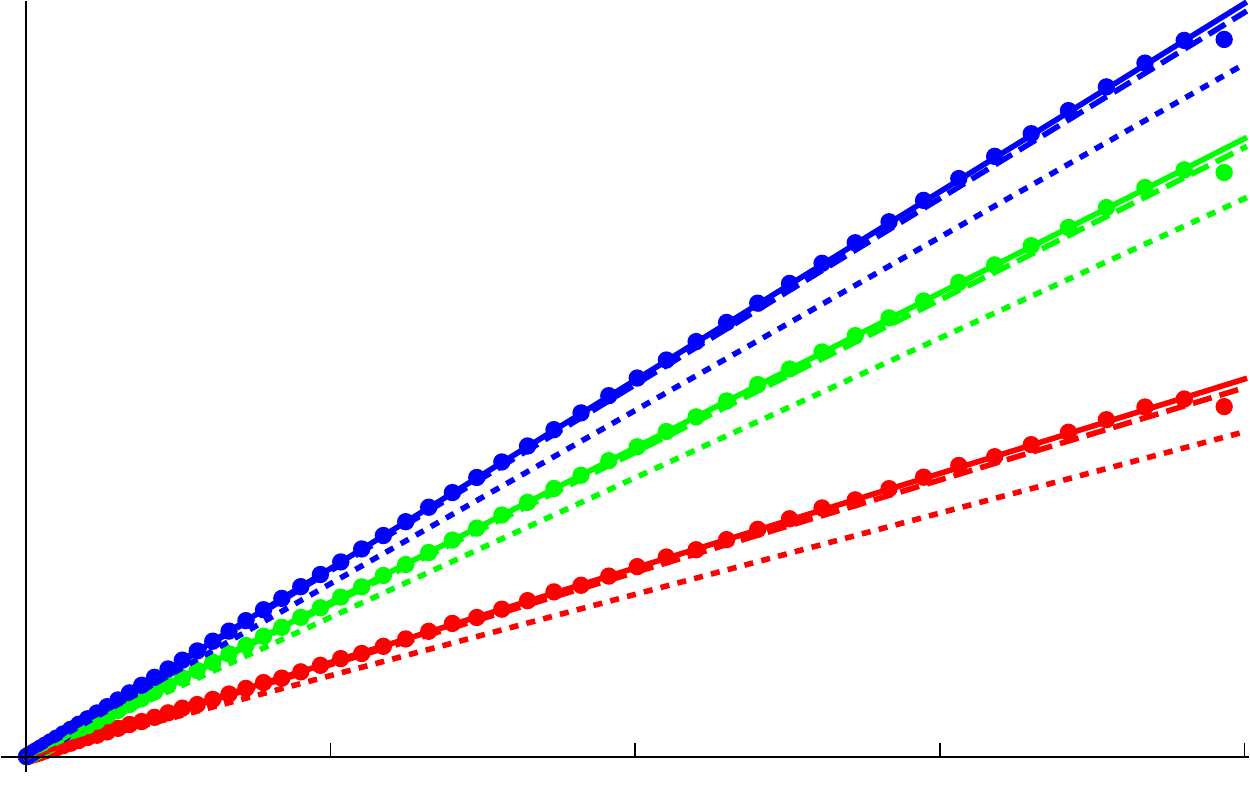}}
\put(57.25,1){$R^2$}
\put(4.5,33.75){$S_{\mathrm{EE}}$}
\put(27.75,-1){$N a / 2$}
\put(54.25,-1){$N a$}
\put(58.5,7.75){\line(0,1){22}}
\put(58.5,7.75){\line(1,0){19}}
\put(77.5,7.75){\line(0,1){22}}
\put(58.5,29.75){\line(1,0){19}}
\put(59,20){\includegraphics[width = 0.07\textwidth]{3dlegend.pdf}}
\put(60.25,16.25){\includegraphics[width = 0.05\textwidth]{3dlegenddot.pdf}}
\put(60,7.75){\includegraphics[width = 0.05125\textwidth]{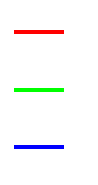}}
\put(65.5,27.125){1st order}
\put(65.5,24.125){2nd order}
\put(65.5,21.125){3rd order}
\put(65.5,18.25){numerical}
\put(65.5,15.125){$c=\sqrt{\pi}$}
\put(65.5,12.125){$c=2\sqrt{\pi}$}
\put(65.5,9.125){$c=4\sqrt{\pi}$}
\end{picture}
\caption{The entanglement entropy in scalar field theory in $3+1$ dimensions with an azimuthal cutoff of the form $\ell_{\max} = c R / a$ for $ma=1$ and various values of the constant $c$}
\label{fig:lmax}
\end{figure}

In order to investigate whether the inverse mass expansion is still a good approximation when an azimuthal cutoff of the form $\ell_{\max} = c R / a$ is introduced, the entanglement entropy in $3+1$ dimensions for $\mu a = 1$ and various values of $c$ is numerically computed and compared to the perturbative formulae \eqref{eq:SEE_c} in figure \ref{fig:lmax}. 

We may conclude the following:
\begin{itemize}
\item An azimuthal cutoff of the form $\ell_{\max} = c R / a$ preserves the dominance of the area law term in entanglement entropy. This is not the case when a more general azimuthal cutoff is chosen (e.g. $\ell_{\max} = c$). The inverse mass expansion is still a good approximation when such a regularization scheme is chosen.
\item The area law term, as well as the subleading terms, are strongly affected by the selection of the dependence of the azimuthal cutoff $\ell_{\max}$ on the radial cutoff $a$. This is the expected behaviour comparing with calculations in CFT or holographic calculations via the Ryu-Takayanagi conjecture. The only terms that do not depend on the regularization scheme are the universal terms, which cannot be captured by our perturbation theory.
\item The introduction of an azimuthal cutoff would also set the perturbative calculation of the entanglement entropy finite at higher number of dimensions, where the respective integral term diverges as $\ell_{\max} \to \infty$.
\item Srednicki's calculation, which is equivalent to the specific choice $c \to \infty$, is an upper bound for the area law coefficient. The fact that the integral terms in more than $3+1$ dimensions diverge, implies that such an upper bound exists only in $2+1$ and $3+1$ dimensions.
\end{itemize}

\section{Discussion}
\label{sec:discussion}

The calculation of entanglement entropy in the ground state of oscillatory systems, which include free scalar field theories, at their ground state is in general a difficult, non-perturbative calculation, since the ground state is highly entangled. We managed to find a perturbative method to calculate it, using as expansive parameter the ratio of the non-diagonal to diagonal elements of the couplings matrix of the system. This parameter in the case of free scalar field theory is being played by the inverse mass of the field. 

The calculation of entanglement entropy in the inverse mass expansion indicates that the major contribution to entanglement entropy is a term proportional to the area of the entangling surface, i.e. the ``area law'' term, a well-known fact since \cite{Srednicki:1993im,Bombelli:1986rw}. The perturbative calculation of the coefficient of this term agrees with the numerical calculation of entanglement entropy, based on the techniques of \cite{Srednicki:1993im}, and provides an analytic method for the specification of such coefficients. Subleading terms in the expansion of entanglement entropy for large entangling sphere radii can also be perturbatively calculated. The inverse mass expansion and the entangling sphere radius expansions can be performed simultaneously, but they are not parallel in any sense. The leading term in the entangling sphere radius expansion, i.e. the area law term, as well as the subleading terms, receive contributions at all orders in the inverse mass expansion.

The area law term, as well as the subleading ones are dependent on the regularization scheme, in line with analogous replica trick calculations. Universal terms that appear in the massless limit and depend on the global characteristics of the entangling surface (logarithmic terms in even dimensions and constant terms in odd dimensions) are non-perturbative contributions in this expansive approach. Furthermore, in this approach, the coefficient of the area law term in $2+1$ and $3+1$ dimensions has an upper bound, for any regularization scheme. The latter does not exist in higher dimensions.

An interesting feature of the inverse mass expansion is the following: the perturbation parameter is not exactly the inverse mass, but rather the quantity $1 / \sqrt{\mu^2 a^2 + 2}$, where $a$ is the UV cutoff length scale imposed in the radial direction. This fact allows the application of the perturbation series even in the massless field case. Not surprisingly, the perturbation series converges more slowly than in the massive case; however, the values of the first terms strongly suggest that it still converges to the numerical results. In the case of free massless scalar field in $3+1$ dimensions the inverse mass series for the coefficient of the area law term approaches the value $0.295$ found in \cite{Srednicki:1993im,Lohmayer:2009sq}.

An important advantage of the presented perturbative method is that it is not limited to the calculation of entanglement entropy, but it provides the full spectrum of the reduced density matrix. The latter, unlike entanglement entropy, contains the full information of the entanglement between the considered subsystems. This is clearly an advantage in comparison to holographic calculations which are constrained to the specification of entanglement entropy due to the nature of the Ryu-Takayanagi conjecture (the latter of course can be applied to strongly coupled system, where it is impossible to apply our perturbative method). As long as calculations based on the replica trick are concerned, they naturally allow the calculation of \ren entropies $S_q$ for all $q$. Although in principle it is possible to reconstruct the spectrum of the reduced density matrix from the latter, in practise this process is very complicated and usually only the specification of the largest eigenvalue and its degeneracy may be easily achieved.

This perturbative method is an appropriate tool to expose the connection between the ``area law'' and the locality of the underlying field theory. Locality is encoded into the couplings matrix $K$ as the absence of non-diagonal elements apart from the elements of the superdiagonal and subdiagonal. This in turn results in an hierarchy for the eigenvalues of the reduced density matrix system, leading to the area law. This hierarchy in the spectrum of the reduced density matrix depicts the fact that locality enforces entanglement between the interior and the exterior of the sphere to be dominated by the entanglement between pairs of neighbouring degrees of freedom that are separated by the entangling surface. The latter are clearly proportional to the area and not the volume of the entangling sphere.

It would be interesting to extend the applications of this perturbative expansion to other geometries, e.g. dS or AdS spacetimes, to cases where the overall system does not lie at its ground state (e.g. systems at a thermal state) or to other field theories containing fermionic fields or gauge fields. Furthermore, application of the above techniques for non-spherical entangling surfaces may shed light to the dependence of entanglement entropy on the geometric features of the latter, such as the curvature.


\subsection*{Acknowledgments}

The research of G.P. is funded by the ``Post-doctoral researchers support'' action of the operational programme ``human resources development, education and long life learning, 2014-2020'', with priority axes 6, 8 and 9, implemented by the Greek State Scholarship Foundation and co-funded by the European Social Fund - ESF and National Resources of Greece.

The authors would like to thank M. Axenides, E. Floratos and G. Linardopoulos for useful discussions.

\appendix
\renewcommand{\thesection}{\Alph{section}}
\renewcommand{\thesubsection}{\Alph{section}.\arabic{subsection}}
\renewcommand{\theequation}{\Alph{section}.\arabic{equation}}

\setcounter{equation}{0}

\section{Discretization of the Scalar Field Theory}
\label{sec:descretization}
\subsubsection*{$3+1$ Dimensions}
We consider a free real scalar field theory in $3+1$ dimensions. The Hamiltonian reads
\begin{equation}
H = \frac{1}{2}\int {{d^3}x\left[ {{\pi ^2}\left( {\vec x} \right) + {{\left| {\vec \nabla \varphi \left( {\vec x} \right)} \right|}^2} + {\mu^2}{\varphi^2} {{\left( {\vec x} \right)}}} \right]} .
\label{eq:discretize_hamiltonian}
\end{equation}
We define,
\begin{align}
{\varphi _{\ell m}}\left( r \right) &= r\int {d\Omega {Y_{\ell m}}\left( {\theta ,\varphi } \right)\varphi \left( {\vec x} \right)} ,\label{eq:discretize_def1}\\
{\pi _{\ell m}}\left( r \right) &= r\int {d\Omega {Y_{\ell m}}\left( {\theta ,\varphi } \right)\pi \left( {\vec x} \right)} ,\label{eq:discretize_def2}
\end{align}
where $r=\left|\vec{x}\right|$ is the radial coordinate and ${Y_{\ell m}}$ are the real spherical harmonics, defined as,
\begin{equation}
{Y_{\ell m}} = \begin{cases}
\sqrt 2 {\left( { - 1} \right)^m}{\mathop{\rm Im}\nolimits} \left[ {Y_\ell^{ - m}} \right], & m<0 ,\\
Y_l^0, & m=0, \\
\sqrt 2 {\left( { - 1} \right)^m}{\mathop{\rm Re}\nolimits} \left[ {Y_\ell^m} \right], & m>0 .
\end{cases}
\end{equation}
The real spherical harmonics form an orthonormal basis of harmonic functions on the sphere $S^2$. It is easy to show that quantities ${\varphi _{\ell m}}\left( r \right)$ and ${\pi _{\ell m}}\left( r \right)$ obey canonical commutation relations,
\begin{equation}
\left[ {{\varphi _{\ell m}}\left( r \right),{\pi _{\ell ' m'}}\left( r' \right)} \right] = i\delta \left( {r - r'} \right){\delta _{\ell \ell '}}{\delta _{mm'}} .
\end{equation}

Expanding the field in real spherical harmonics and substituting in \eqref{eq:discretize_hamiltonian}, we acquire an expression of the Hamiltonian in terms of ${\varphi _{\ell m}}\left( r \right)$ and ${\pi _{\ell m}}\left( r \right)$,
\begin{equation}
H = \frac{1}{2}\sum\limits_{\ell ,m} {\int_0^\infty  {dr \left\{ {\pi _{\ell m}^2\left( r \right) + {r^2}{{\left[ {\frac{\partial }{{\partial r}}\left( {\frac{{{\varphi _{\ell m}}\left( r \right)}}{r}} \right)} \right]}^2} + \left( {\frac{{\ell \left( {\ell  + 1} \right)}}{{{r^2}}} + {\mu^2}} \right)\varphi _{\ell m}^2\left( r \right)} \right\}} } .
\label{eq:free_massive_continuous}
\end{equation}

The only continuous variable left is the radial coordinate $r$. We regularize the theory introducing a lattice of $N$ spherical shells with radii $r_i = i a$ with $i \in \mathds{N}$ and $1 \le i \le N$. The Hamiltonian of the discretized system can be found via the application of the following rules on equation {\eqref{eq:free_massive_continuous}:
\begin{equation}
\begin{split}
r &\to ja\\
{\varphi _{\ell m}}\left( {ja} \right) &\to {\varphi _{\ell m,j}} , \\
\left. \frac{{\partial {\varphi _{\ell m}}\left( r \right)}}{{\partial r}} \right|_{r=ja} &\to \frac{{{\varphi _{\ell m,j + 1}} - {\varphi _{\ell m,j}}}}{a} , \\
{\pi _{\ell m}}\left( {ja} \right) &\to \frac{{{\pi _{\ell m,j}}}}{a} , \\
\int_0^{\left( {N + 1} \right)a} {dr}  &\to a\sum\limits_{j = 1}^N {} .
\end{split}
\label{eq:discretization_scheme}
\end{equation}
The discretized Hamiltonian reads
\begin{equation}
H = \frac{1}{{2a}}\sum\limits_{\ell ,m} {\sum\limits_{j = 1}^N {\left[ {\pi _{\ell m,j}^2 + {{\left( {j + \frac{1}{2}} \right)}^2}{{\left( {\frac{{{\varphi _{\ell m,j + 1}}}}{{j + 1}} - \frac{{{\varphi _{\ell m,j}}}}{j}} \right)}^2} + \left( {\frac{{\ell \left( {\ell  + 1} \right)}}{{{j^2}}} + {\mu^2}{a^2}} \right)\varphi _{\ell m,j}^2} \right]} } .
\end{equation}

\subsubsection*{$2+1$ Dimensions}

We may study free scalar field theory in $2+1$ dimensions in a similar manner. The Hamiltonian reads
\begin{equation}
H = \frac{1}{2}\int {{d^2}x\left[ {{\pi ^2}\left( {\vec x} \right) + {{\left| {\vec \nabla \varphi \left( {\vec x} \right)} \right|}^2} + {\mu^2}{\varphi^2} {{\left( {\vec x} \right)}}} \right]} .
\label{eq:discretize_hamiltonian_d2}
\end{equation}
We define,
\begin{align}
{\varphi _\ell}\left( r \right) &= \sqrt{r}\int {d\theta {Y_\ell}\left( {\theta} \right)\varphi \left( {\vec x} \right)} ,\label{eq:discretize_def1_d2}\\
{\pi _\ell}\left( r \right) &= \sqrt{r}\int {d\theta {Y_\ell}\left( {\theta} \right)\pi \left( {\vec x} \right)} ,\label{eq:discretize_def2_d2}
\end{align}
where $r$ is the radial coordinate and ${Y_\ell}$ are the real circular harmonics,
\begin{equation}
{Y_\ell} = \begin{cases}
{\sin \left( \ell \theta \right)}/{\sqrt{ \pi }}, & \ell<0 ,\\
{1}/{\sqrt{2\pi}} & \ell=0, \\
{\cos \left( \ell \theta \right)}/{\sqrt{ \pi }}, & \ell>0 .
\end{cases}
\end{equation}
The functions ${Y_\ell}$ form an orthonormal basis of harmonic functions on the circle $S^1$. The quantities ${\varphi _{\ell}}\left( r \right)$ and ${\pi _{\ell}}\left( r \right)$ obey canonical commutation relations,
\begin{equation}
\left[ {{\varphi _{\ell}}\left( r \right),{\pi _{\ell '}}\left( r' \right)} \right] = i\delta \left( {r - r'} \right){\delta _{\ell \ell '}} .
\end{equation}

We expand the field in real circular harmonics and substitute in \eqref{eq:discretize_hamiltonian_d2} to find
\begin{equation}
H = \frac{1}{2}\sum\limits_{\ell } {\int_0^\infty  {dr \left\{ {\pi _{\ell}^2\left( r \right) + {r}{{\left[ {\frac{\partial }{{\partial r}}\left( {\frac{{{\varphi _{\ell}}\left( r \right)}}{\sqrt{r}}} \right)} \right]}^2} + \left( {\frac{{\ell^2 }}{{{r^2}}} + {\mu^2}} \right)\varphi _{\ell }^2\left( r \right)} \right\}} } .
\label{eq:free_massive_continuous_d2}
\end{equation}
Using the discretization scheme \eqref{eq:discretization_scheme}, we obtain the discretized Hamiltonian
\begin{equation}
H = \frac{1}{{2a}}\sum\limits_{\ell} {\sum\limits_{j = 1}^N {\left[ {\pi _{\ell,j}^2 + {{\left( {j + \frac{1}{2}} \right)}}{{\left( {\frac{{{\varphi _{\ell ,j + 1}}}}{{\sqrt{j + 1}}} - \frac{{{\varphi _{\ell ,j}}}}{\sqrt{j}}} \right)}^2} + \left( {\frac{{\ell^2 }}{{{j^2}}} + {\mu^2}{a^2}} \right)\varphi _{\ell ,j}^2} \right]} } .
\end{equation}

\section{The Inverse Mass Expansion at Second and Third Order}
\label{sec:third}
It is not difficult to show that there are no corrections to the entanglement entropy at the next to leading order in $\varepsilon$. Therefore, the first corrections appear at third order in the inverse mass expansion. For this purpose it is required that the matrix $\Omega$ is calculated with four non-vanishing terms.
Following the definitions \eqref{eq:K_ij_def}, \eqref{eq:k_i_def} and \eqref{eq:l_i_def}, we find that the matrix $\Omega$ at order $\varepsilon^5$ is given by
\begin{multline}
{\Omega _{ij}} = \left( {\omega _i^{0\left( { - 1} \right)}{\varepsilon ^{ - 1}} + \omega _i^{0\left( 3 \right)}{\varepsilon ^3}} \right){\delta _{ij}} + \left( {\omega _i^{1\left( 1 \right)}{\varepsilon ^1} + \omega _i^{1\left( 5 \right)}{\varepsilon ^5}} \right)\left( {{\delta _{i + 1,j}} + {\delta _{i,j + 1}}} \right)\\
 + \omega _i^{2\left( 3 \right)}{\varepsilon ^3}\left( {{\delta _{i + 2,j}} + {\delta _{i,j + 2}}} \right) + \omega _i^{3\left( 5 \right)}{\varepsilon ^5}\left( {{\delta _{i + 3,j}} + {\delta _{i,j + 3}}} \right) + \mathcal{O}\left( {{\varepsilon ^7}} \right) ,
\end{multline}
where
\begin{align}
&\omega _i^{0\left( { - 1} \right)} = {k_i} , \\
&\omega _i^{0\left( 3 \right)} =  - \frac{1}{{2{k_i}}}\left( {l_i^2\left( {1 - {\delta _{iN}}} \right) + l_{i - 1}^2\left( {1 - {\delta _{i1}}} \right)} \right) , \\
&\omega _i^{1\left( 1 \right)} = {l_i} , \\
&\omega _i^{1\left( 5 \right)} = \frac{1}{2}\frac{{{l_i}}}{{{k_i} + {k_{i + 1}}}}\left[ {l_i^2\left( {\frac{1}{{{k_i}}} + \frac{1}{{{k_{i + 1}}}}} \right)} \right. \nonumber \\
&\left. { + l_{i - 1}^2\left( {1 - {\delta _{i1}}} \right)\left( {\frac{1}{{{k_i}}} + \frac{2}{{{k_{i - 1}} + {k_{i + 1}}}}} \right) + l_{i + 1}^2\left( {1 - {\delta _{i,N - 1}}} \right)\left( {\frac{1}{{{k_{i + 1}}}} + \frac{2}{{{k_i} + {k_{i + 2}}}}} \right)} \right] , \label{eq:edge_effect_omega}\\
&\omega _i^{2\left( 3 \right)} =  - \frac{{{l_i}{l_{i + 1}}}}{{{k_i} + {k_{i + 2}}}} , \\
&\omega _i^{3\left( 5 \right)} = \frac{{{l_i}{l_{i + 1}}{l_{i + 2}}\left( {{k_i} + {k_{i + 1}} + {k_{i + 3}} + {k_{i + 3}}} \right)}}{{\left( {{k_i} + {k_{i + 2}}} \right)\left( {{k_{i + 1}} + {k_{i + 3}}} \right)\left( {{k_i} + {k_{i + 3}}} \right)}} .
\end{align}

Trivially,
\begin{align}
{A_{ij}} &= {\Omega _{ij}},\quad i = 1, \ldots ,n,\quad j = 1, \ldots ,n,\\
{B_{ij}} &= {\Omega _{i,j + n}},\quad i = 1, \ldots ,n,\quad j = 1, \ldots ,N - n,\\
{C_{ij}} &= {\Omega _{i + n,j + n}},\quad i = 1, \ldots ,N - n,\quad j = 1, \ldots ,N - n.
\end{align}
The matrix $B$ has only a finite set of elements not vanishing at this order, namely,
\begin{multline}
{B_{ij}} = \left( {\omega _n^{1\left( 1 \right)}\varepsilon  + \omega _n^{1\left( 5 \right)}{\varepsilon ^5}} \right){\delta _{in}}{\delta _{j1}} + \omega _{n - 1}^{2\left( 3 \right)}{\varepsilon ^3}{\delta _{i,n - 1}}{\delta _{j1}} + \omega _n^{2\left( 3 \right)}{\varepsilon ^3}{\delta _{in}}{\delta _{j2}}\\
 + \omega _{n - 2}^{3\left( 5 \right)}{\varepsilon ^5}{\delta _{i,n - 2}}{\delta _{j1}} + \omega _{n - 1}^{3\left( 5 \right)}{\varepsilon ^5}{\delta _{i,n - 1}}{\delta _{j2}} + \omega _n^{3\left( 5 \right)}{\varepsilon ^5}{\delta _{in}}{\delta _{j3}} + \mathcal{O}\left( {{\varepsilon ^7}} \right) .
\end{multline}

We need to acquire the matrices $A^{-1}$ and $C^{-1}$ with three non-vanishing terms. They equal
\begin{multline}
{\left( {{A^{ - 1}}} \right)_{ij}} = \left( {a_i^{0\left( 1 \right)}{\varepsilon } + a_i^{0\left( 5 \right)}{\varepsilon ^5}} \right){\delta _{ij}} + a_i^{1\left( 3 \right)}{\varepsilon ^3}\left( {{\delta _{i + 1,j}} + {\delta _{i,j + 1}}} \right) \\
+ a_i^{2\left( 5 \right)}{\varepsilon ^5}\left( {{\delta _{i + 2,j}} + {\delta _{i,j + 2}}} \right) + \mathcal{O}\left( {{\varepsilon ^7}} \right) ,
\end{multline}
where
\begin{align}
a_i^{0\left( 1 \right)} &= \frac{1}{{{k_i}}} , \\
a_i^{0\left( 5 \right)} &= \frac{1}{{k_i^2}}\left[ {l_{i - 1}^2\left( {1 - {\delta _{i1}}} \right)\left( {\frac{1}{{{k_{i - 1}}}} + \frac{1}{{2{k_i}}}} \right) + l_i^2\left( {\frac{{1 - {\delta _{in}}}}{{{k_{i + 1}}}} + \frac{1}{{2{k_i}}}} \right)} \right] , \label{eq:edge_effect_2}\\
a_i^{1\left( 3 \right)} &= - \frac{{{l_i}}}{{{k_i}{k_{i + 1}}}} , \\
a_i^{2\left( 5 \right)} &= \frac{{{l_i}{l_{i + 1}}}}{{{k_i}{k_{i + 2}}}}\left( {\frac{1}{{{k_{i + 1}}}} + \frac{1}{{{k_i} + {k_{i + 2}}}}} \right) .
\end{align}
Similarly,
\begin{multline}
{\left( {{C^{ - 1}}} \right)_{ij}} = \left( {c_i^{0\left( 1 \right)}\varepsilon  + c_i^{0\left( 5 \right)}{\varepsilon ^5}} \right){\delta _{ij}} + c_i^{1\left( 3 \right)}{\varepsilon ^3}\left( {{\delta _{i + 1,j}} + {\delta _{i,j + 1}}} \right)\\
 + c_i^{2\left( 5 \right)}{\varepsilon ^5}\left( {{\delta _{i + 2,j}} + {\delta _{i,j + 2}}} \right) + \mathcal{O}\left( {{\varepsilon ^7}} \right) ,
\end{multline}
where
\begin{align}
c_i^{0\left( 1 \right)} &= \frac{1}{{{k_{i + n}}}} , \\
c_i^{0\left( 5 \right)} &= \frac{1}{{k_{i + n}^2}} \left[ l_{i + n}^2\left( {1 - {\delta _{i,N - n}}} \right)\left( {\frac{1}{{{k_{i + n + 1}}}} + \frac{1}{{2{k_{i + n}}}}} \right) + l_{i + n - 1}^2\left( {\frac{{1 - {\delta _{i1}}}}{{{k_{i + n - 1}}}} + \frac{1}{{2{k_{i + n}}}}} \right) \right] , \\
c_i^{1\left( 3 \right)} &= - \frac{{{l_{i + n}}}}{{{k_{i + n}}{k_{i + n + 1}}}} , \\
c_i^{2\left( 5 \right)} &= \frac{{{l_{i + n}}{l_{i + n + 1}}}}{{{k_{i + n}}{k_{i + n + 2}}}}\left( {\frac{1}{{{k_{i + n + 1}}}} + \frac{1}{{{k_{i + n + 2}} + {k_{i + n}}}}} \right) .
\end{align}

It is a matter of algebra to show that the matrix ${{\gamma ^{ - 1}}\beta }$ has a finite number of non-vanishing elements at this order, namely,
\begin{multline}
{\left( {{\gamma ^{ - 1}}\beta } \right)_{ij}} = \left( {\beta _{11}^{\left( 4 \right)}{\varepsilon^4} + \beta _{11}^{\left( 8 \right)}{\varepsilon^8}} \right){\delta _{i1}}{\delta _{j1}} + \beta _{21}^{\left( 6 \right)}{\varepsilon^6}{\delta _{i2}}{\delta _{j1}} + \beta _{12}^{\left( 6 \right)}{\varepsilon^6}{\delta _{i1}}{\delta _{j2}}\\
 + \beta _{31}^{\left( 8 \right)}{\varepsilon^8}{\delta _{i3}}{\delta _{j1}} + \beta _{13}^{\left( 8 \right)}{\varepsilon^8}{\delta _{i1}}{\delta _{j3}} + \beta _{22}^{\left( 8 \right)}{\varepsilon^8}{\delta _{i2}}{\delta _{j2}} + \mathcal{O}\left( {{\varepsilon^{10}}} \right) .
\end{multline}

Up to this order, the matrix ${{\gamma ^{ - 1}}\beta }$ has in general two non-vanishing eigenvalues
\begin{align}
{\lambda _{1}} &= \beta _{11}^{\left( 4 \right)}{\varepsilon^4} + \left( {\beta _{11}^{\left( 8 \right)} + \frac{{\beta _{12}^{\left( 6 \right)}\beta _{21}^{\left( 6 \right)}}}{{\beta _{11}^{\left( 4 \right)}}}} \right){\varepsilon^8} + \mathcal{O}\left(\varepsilon^{12}\right),\\
{\lambda _{2}} &= \left( {\beta _{22}^{\left( 8 \right)} - \frac{{\beta _{12}^{\left( 6 \right)}\beta _{21}^{\left( 6 \right)}}}{{\beta _{11}^{\left( 4 \right)}}}} \right){\varepsilon^8} + \mathcal{O}\left(\varepsilon^{12}\right) .
\end{align}
The eigenvalue ${\lambda _{2}}$ turns out to vanish at this order, whereas
\begin{multline}
{\lambda _{1}} = \frac{{l_n^2}}{{2{k_n}{k_{n + 1}}}}{\varepsilon ^4} + \frac{{l_n^2}}{{2{k_n}{k_{n + 1}}}}\Bigg[ {\frac{{l_n^2}}{2}\left( {{{\left( {\frac{1}{{{k_n}}} + \frac{1}{{{k_{n + 1}}}}} \right)}^2} + \frac{1}{{{k_n}{k_{n + 1}}}}} \right)} \\
 + l_{n + 1}^2\left( {\frac{1}{{2k_{n + 1}^2}} + \frac{{{k_{n + 1}}}}{{{k_{n + 2}}{{\left( {{k_n} + {k_{n + 2}}} \right)}^2}}} + \frac{{\left( {{k_n} + {k_{n + 1}} + {k_{n + 2}}} \right)\left( {{k_n} + 2{k_{n + 1}} + {k_{n + 2}}} \right)}}{{{k_{n + 1}}{k_{n + 2}}\left( {{k_n} + {k_{n + 1}}} \right)\left( {{k_n} + {k_{n + 2}}} \right)}}} \right)\\
{ + l_{n - 1}^2\left( {\frac{1}{{2k_n^2}} + \frac{{{k_n}}}{{{k_{n - 1}}{{\left( {{k_{n - 1}} + {k_{n + 1}}} \right)}^2}}} + \frac{{\left( {{k_{n - 1}} + {k_n} + {k_{n + 1}}} \right)\left( {{k_{n - 1}} + 2{k_n} + {k_{n + 1}}} \right)}}{{{k_{n - 1}}{k_n}\left( {{k_n} + {k_{n + 1}}} \right)\left( {{k_{n - 1}} + {k_{n + 1}}} \right)}}} \right)} \Bigg]{\varepsilon ^8} .
\label{eq:eigenvalue_2nd_order}
\end{multline}

At third non-vanishing order in the inverse mass expansion, another non-vanishing eigenvalue emerges having a leading contribution of order $\varepsilon ^{12}$. Considering only the area law term contribution to entanglement entropy is equivalent to approximating all $k_i$ and $l_i$ with $k_{n_R}$ and $-1$ respectively. At this approximation and without showing more details, the eigenvalues of ${{\gamma ^{ - 1}}\beta }$ at third order in the inverse mass expansion equal
\begin{align}
{\lambda _{1}} &= \frac{1}{{8k_{n_R}^4}} + \frac{5}{{16k_{n_R}^8}} + \frac{1875}{{2048 k_{n_R}^{12}}} + \mathcal{O} \left( \mu^{-16} \right) , \label{eq:eigenvalue_3rd_1}\\
{\lambda _{2}} &= \frac{1}{{2048 k_{n_R}^{12}}} + \mathcal{O} \left( \mu^{-16} \right) . \label{eq:eigenvalue_3rd_2}
\end{align}

The entanglement entropy at this order reads
\begin{multline}
{S_{{\rm{EE}}\ell }} = \frac{{\lambda _{1}^{\left( 4 \right)}}}{2}\left( {1 - \ln \frac{{\lambda _{1}^{\left( 4 \right)}{\varepsilon ^4}}}{2}} \right){\varepsilon ^4} + \left[ {\frac{{{{\left( {\lambda _{1}^{\left( 4 \right)}} \right)}^2}}}{8}\left( {1 - 2 \ln \frac{{\lambda _{1}^{\left( 4 \right)}{\varepsilon ^4}}}{2}} \right) - \frac{{\lambda _{1}^{\left( 8 \right)}}}{2}\ln \frac{{\lambda _{1}^{\left( 4 \right)}{\varepsilon ^4}}}{2}} \right]{\varepsilon ^8} \\
+ \left[ {\frac{{{{\left( {\lambda _1^{\left( 4 \right)}} \right)}^3}}}{{24}}\left( {1 - 6\ln \frac{{\lambda _1^{\left( 4 \right)}{\varepsilon ^4}}}{2}} \right) - \frac{{{{\left( {\lambda _1^{\left( 8 \right)}} \right)}^2}}}{{4\lambda _1^{\left( 4 \right)}}} - \frac{{\lambda _1^{\left( 4 \right)}\lambda _1^{\left( 8 \right)} + \lambda _1^{\left( {12} \right)}}}{2}\ln \frac{{\lambda _1^{\left( 4 \right)}{\varepsilon ^4}}}{2} } \right. \\
\left. {\phantom{{\frac{{{{\left( {\lambda _1^{\left( 4 \right)}} \right)}^3}}}{{24}}}} + \frac{{\lambda _2^{\left( {12} \right)}}}{2}\left( {1 - \ln \frac{{\lambda _2^{\left( {12} \right)}{\varepsilon ^{12}}}}{2}} \right)} \right]{\varepsilon ^{12}} + \mathcal{O}\left( {{\varepsilon ^{16}}} \right) ,
\label{eq:SEEl_second_order}
\end{multline}
where ${\lambda _{1}^{\left( 4,8,12 \right)}}$ and ${\lambda _{2}^{\left( 12 \right)}}$ may be read from equations \eqref{eq:eigenvalue_2nd_order}, \eqref{eq:eigenvalue_3rd_1} and \eqref{eq:eigenvalue_3rd_2}. 

\section{Numerical Calculation Code}
\label{sec:code}

In section \ref{sec:area}, the perturbative formulae for entanglement entropy were compared with a numerical calculation of the latter. The numerical algorithm uses the same regularization scheme as the perturbative expansion, but it calculates the matrices $\beta$, $\gamma$, as well as the eigenvalues of the matrix $\gamma^{-1} \beta$, numerically. The numerical calculation was performed with the help of the following code in Wolfram's Mathematica.

\begin{verbatim}
xi[beta_] := beta / (1 + Sqrt[1 - beta^2]);
S[xi_] := - Log[1 - xi] - xi / (1 - xi) * Log[xi] /; xi > 0;
S[xi_] := 0 /; xi <= 0; (*to prevent errors from vanishing eigenvalues*)
mass_sq = 1;
Nmax = 60;
elmax = 1000;
entropies_el = Table[0, {el, 1, elmax + 1}, {n, 1, Nmax - 1}];
ent_entropy = Table[0, {n, 1, Nmax - 1}];

For[el = 0, el < elmax + 1, el++, 
  K = Table[
    KroneckerDelta[i,j]*((j + 1/2)^2/j^2 
    + (j - 1/2)^2 / j^2 * HeavisideTheta[j - 3/2] 
    + (el * (el + 1)) / j^2 + mass_sq) - 
    KroneckerDelta[i, j + 1] (j + 1/2)^2 / (j * (j + 1)) - 
    KroneckerDelta[i + 1, j] (i + 1/2)^2 / (i * (i + 1)), 
    {i, 1, Nmax}, {j, 1, Nmax}
  ];
  Omega = MatrixPower[N[K], 1/2];
  For[n = 1, n < Nmax, n++,
    KA = Omega[[1 ;; n, 1 ;; n]];
    KB = Omega[[1 ;; n, n + 1 ;; Nmax]];
    KC = Omega[[n + 1 ;; Nmax, n + 1 ;; Nmax]];
    beta = (1/2) * Transpose[KB].Inverse[KA].KB;
    gamma = KC - beta;
    lambdas = Eigenvalues[Inverse[gamma].beta];
    entropies_el[[el + 1, n]] =
    Sum[S[xi[lambdas[[i]]]], {i, 1, Nmax - n}];
    (*el=0 contribution is written in the entropies_el[[1]] element*)
  ];
]
For[n = 1, n < Nmax, n++, 
 ent_entropy[[n]] = 
 Sum[(2 * (el - 1) + 1) * entropies_el[[el, n]], {el, 1, elmax + 1}]]
\end{verbatim}

The above code applies in the case of $3+1$ dimensions. For the numerical calculation it is necessary to impose a cutoff $\ell_{\max}$ to the values of $\ell$. This has been selected appropriately large so that the series has converged close enough to the $\ell_{\max} \to \infty$ limit (such an appropriate choice of $\ell_{\max}$ depends on the value of the mass parameter). Specification of terms beyond the area law term requires much larger values of $\ell_{\max}$, and, thus, running time. The required modifications for the numerical calculation of entanglement entropy in different number of dimensions or the introduction of an angular cutoff that depends on the entangling sphere radius are quite simple.

\end{document}